\documentclass[%
reprint,
superscriptaddress,
nofootinbib,
amsmath,amssymb,
aps,
onecolumn,
]{revtex4-2}
\UseRawInputEncoding
\usepackage{float}
\usepackage{dsfont}
\usepackage{physics}
\usepackage{mathrsfs}
\usepackage{amsmath}
\usepackage{indentfirst}
\usepackage{amssymb}
\usepackage{color}
\usepackage{amsfonts} 
\usepackage[framemethod=TikZ]{mdframed}
\usepackage{mathtools}
\usepackage[retainorgcmds]{IEEEtrantools}
\usepackage[colorlinks=true,linkcolor=blue,urlcolor=blue,citecolor=blue,pdfusetitle]{hyperref}
\usepackage{times,txfonts}
\usepackage{graphicx}
\usepackage{dcolumn}
\usepackage{bm}
\usepackage{tabularx}

\begin{document}

\title{FRACTIONAL CONTROL GATE PROTOCOLS FOR QUANTUM ENGINES}
\author{Elliot Fox}
\email{elliot.fox@york.ac.uk}
\affiliation{School of Physics, Engineering and Technology, University of York, York YO10 5DD, United Kingdom}
\author{Taysa Mendes de Mendon\c{c}a}
\email{tmendonca@ifsc.usp.br}
\affiliation{Instituto de F\'isica de S\~ao Carlos, Universidade de S\~ao Paulo, IFSC – USP, 13566-590, S\~ao Carlos, SP, Brazil.}
\author{Ferdinand Schmidt-Kaler}
\email{fsk@uni-mainz.de}
\affiliation{Quantum, Institut f\"{u}r Physik, Universit\"{a}t Mainz, D-55128 Mainz, Germany}
\author{Irene D'Amico}
\email{irene.damico@york.ac.uk}
\affiliation{School of Physics, Engineering and Technology, University of York, York YO10 5DD, United Kingdom}
\date{\today}

\begin{abstract}Nth-root gates allow for a paced application of two-qubit operations. We apply them in quantum thermodynamic protocols for operating a quantum heat engine. A set of circuits for two and three qubits are compared by considering maximum work production and related efficiency. Our results show that for all circuits considered and most regions of initial parameter space, quantum coherence of one of the qubits strongly increases the maximum work production and improves the system's performance as a quantum heat engine. In such circuits, coherence is initially imprinted into one of the qubits, improving the overall maximum extractable work. Work gets generated with $84\%$ to $100\%$ efficiency. Further, we uncover a strong linear correlation between work production and many-body correlations in the working medium generated by these gates.
\end{abstract}

\maketitle

\section{Introduction}

Quantum Thermodynamics (QTD) is a relatively young field which over recent years has seen rapid development, becoming an area in which to discuss and challenge well established concepts, such as the laws of thermodynamics at the quantum scale~\cite{thermodynamics_in_the_quantum_regime,deffner2019quantumthermodynamicsintroductionthermodynamics,e15062100,Goold_2016,Vinjanampathy2016,doi:10.1073/pnas.1411728112,PhysRevE.60.2721,PhysRevLett.78.2690,RevModPhys.81.1665,RevModPhys.83.771,PhysRevLett.127.030602,PRXQuantum.3.020101}. This fast development, coupled with technological advancements in the construction and manipulation of microscopic systems, has facilitated the development of experimental thermal engines which operate outside of the thermodynamic limit~\cite{Rosnagel2016,Peterson2019,Klatzow2019,vonLindenfels2019,VanHorne2020,Bouton2021,PhysRevA.106.022436,Herrera2023,uusnäkki2025experimentalrealizationquantumheat,PhysRevLett.128.090602,Zhang_2022}. Quantum effects, such as coherence and quantum correlations, become increasingly important when trying to accurately describe and model these systems showing promise for providing advantages over classical systems in thermodynamic processes~\cite{Herrera2023,Dillenschneider2009,Rossnagel2014,Vinjanampathy2016,Bera2017,Jaramillo_2016}. 

In quantum thermodynamics, quantum resources can be used to cool or heat systems using information processing. The realisation of Quantum Heat Engines (QHEs) demonstrates an application of quantum mechanics for work producing thermal engines in novel scenarios and theoretical proposals are implemented across many different models including but not limited to, quantum computational protocols~\cite{PhysRevE.100.012109}, laser or photocell QHEs~\cite{doi:10.1073/pnas.1110234108,Oh_2020}, magnetically driven QHEs~\cite{PhysRevE.89.052107}, and quantum mechanical adaptations of the Otto cycle~\cite{10.1140/epjd/e2013-40536-0,Rossnagel2014}. The situation where QHEs outperform their classical counterparts~\cite{Mukherjee_2020,PhysRevLett.128.180602,Jaramillo_2016} or when quantum resources, such as coherence, correlation, or collective behaviour characterised by many-body interactions provide an advantage~\cite{Herrera2023,PhysRevLett.128.180602,PhysRevResearch.4.L032034,Solfanelli_2023,Jaramillo_2016,Hammam_2022} are of particular interest. Hence, considering experimental engines facilitating many-body interactions for these new technologies is of significant importance. Quantum computing-inspired protocols for quantum thermal engines have been proposed, with the experimental realisation of many-body systems implementing quantum logic gates achieved utilising platforms such as, trapped ions~\cite{vonLindenfels2019, PhysRevLett.128.110601}, Rydberg atoms~\cite{Wu2021}, nuclear magnetic resonance~\cite{deAssis2019}, and superconducting qubits~\cite{Abdelhafez2020}. For example, a genuine quantum advantage can be achieved on a quantum processor with correlations between components of the working medium, boosting the extractable work per cycle and producing an efficiency higher than the Carnot standard limit~\cite{Herrera2023}.

In this work, we introduce an algorithmic QHE composed of a three-qubit working medium whose energy dynamics are driven through the stepped application of quantum computational protocols, Figure~\ref{fig:Thermal Machines}. Here, extractable work is characterised as a reduction in energy of the total system with respect to its initial state after the protocol is applied. The problem of practically accessing the extractable work in many-body systems for QHEs in proposed theoretical models remains a challenge. Experimentally, we now have the building blocks to realise a QHE and as the number of qubits available in quantum computers increases, the size of realisable QHEs can increase in turn. The challenge of accessing the extractable work is ongoing, being crucial to the development of a working quantum heat engine. In the present case, the state of our three-qubit system can be accessed through full state quantum tomography~\cite{Blume-Kohout_2010,RevModPhys.81.299} which opens the door for the probing and testing of our proposed engine.

We investigate Nth-root controlled-not logic gates (NRCGs), which when applied iteratively act as a trotterization of controlled-not logic gates, for the implementation of a QHE. NRCGs share similarities with collision models used in other works involving the operation of a QHEs~\cite{Molitor_2020,Bouton2021}. Further, exchanging full Pauli gate operations with a fractional iterative protocol has allowed for investigating the presence of quantum friction, i.e., the generation of quantum coherence in the instantaneous energy eigenbasis under a non-permuting protocol~\cite{Miller2019,Scandi2020}. The~experimental realization of this protocol has proven that quantum friction induces a violation of the work fluctuation dissipation relation, certifying an additional genuine quantum effect~\cite{Onishenko2024}.

The NRCG stepped approach to the dynamics allows access to intermediate work producing states, that, when using full controlled-not logic gates, are typically unavailable, allowing for greater control over the quantity of work produced by the engine. We propose and investigate protocols based on four different circuits. When initial states have quantum coherences, we find that this leads to an increase in maximum work production and a larger work producing parameter region, as compared to initially thermalised qubits. Further, we identify specific regions of heat engine operation and show that when the system is operating in its maximum work producing regime, there are high values of efficiency, demonstrating the suitability of this approach when considering a quantum heat engine. Finally, a strong linear correlation, quantified by the Pearson correlation coefficient, is uncovered between work production and the quantum mutual information  between components of the working medium, generated solely through the application of the NRCG protocols. 

\section{Design and Functioning of the NRCG driven Quantum Heat Engine}\label{theoreticaloutline}

\begin{figure}[hbt!]
    \centering
    \includegraphics[width=15.5cm]{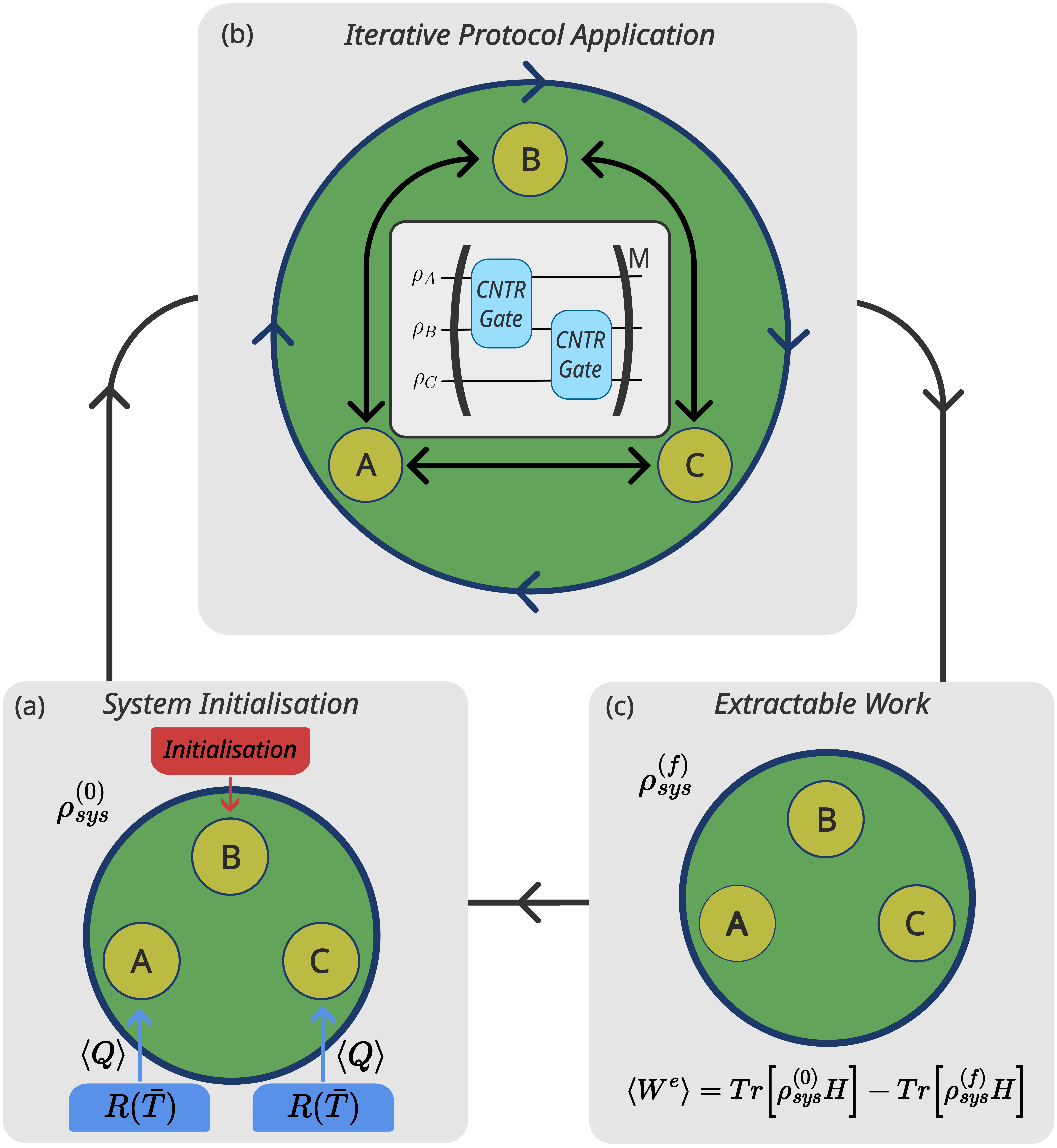}
    \caption{Target operational cycle for the three-qubit quantum heat engine, the two-qubit system is recovered with the removal of qubit-C. The operation is cyclic, where after panel \textbf{(c)} the system can be reinitialised back to its initial state. To note, in this work we are interested only in the dynamics of the sequence \textbf{(a)}, \textbf{(b)}, and \textbf{(c)}. Three distinct steps are shown: \textbf{(a)}: System initialisation - Sec. \ref{Initialisationstep}; \textbf{(b)}: Iterative protocol application - Sec. \ref{Protocol Application Step}; \textbf{(c)}: Extractable work. The arrows shown in panel \textbf{(a)} between the qubits and reservoirs $R(\bar{T})$ indicate positive heat flow $\langle Q \rangle$; the black arrows in \textbf{(b)} show the building of correlations between the qubits mediated by the NRCGs. Except where otherwise stated, we initialise qubit-B in a pure state with initial coherence. Qubits A and C are initialised in identical Gibbs states using the thermal reservoirs $R(\bar{T})$.}
    \label{fig:Thermal Machines}
\end{figure}

We consider a system composed either by two  (A and B) or by three (A, B, and~C) qubits that interact with each other through the use of NRCGs (Figure~\ref{fig:Circuittwobody}). Our system will undergo one operational cycle that consists of three distinct steps: initialisation; iterative protocol application; work extraction. Operations involving logic gates may generate entanglement between the component qubits. Figure~\ref{fig:Thermal Machines} introduces the target operational cycle of our system as a thermal engine, next we will describe in detail each step shown.

\subsection{Initialisation}\label{Initialisationstep}

First, we describe the initialisation step. The component qubits of the working medium are coupled to their respective reservoirs which initialise or re-initialise them to some initial state. Qubits A and C are initialised in identical thermal states at equal temperatures $\bar{T} > 0$ by coupling to the thermal reservoirs $R(\bar{T})$. In the case where qubit-B is initialised in a thermal state, \textit{Initialisation} in Figure~\ref{fig:Thermal Machines}(a) is performed by the reservoir $R(T_B)$. It is important to note there is no initial correlation between any two qubits. The single-qubit thermal state is a Gibbs state defined as~\cite{Lenard1978}

\begin{equation}
 \rho_{Gibbs j} \ = \ (1/Z_{j})\exp^{-\beta_{j} H_{j}},
 \label{eq:gibbsequation}
\end{equation}
where $ Z_{j} = \sum_{i}e^{-\beta_{j} \epsilon_{ij}}$ is the partition function, $\beta_j = (k_BT_j)^{-1}$ is the inverse
temperature parameter, $k_B$ is the Boltzmann constant, and $\epsilon_{i}$ are the eigenvalues of the single-qubit Hamiltonian. This is given by
\begin{equation}
  H_j \ = \
\begin{pmatrix}
\epsilon_{1} & 0\\
0 & \epsilon_{2}
\end{pmatrix}.
\label{H_j}
\end{equation}
As~reference systems, we consider the ones with each qubit prepared in a thermal state: here, there are no initial quantum coherences. We compare these with the systems in which qubit-B is prepared in a pure state; then, initial quantum coherence will transfer through the qubits when a circuit is applied. This allows for probing how initial quantum coherences affect the system's capabilities as a heat engine. Qubit-B is initialised in a pure state as $\rho_{Pure} = \left|\psi\rangle \langle \psi  \right|$, where $|\psi \rangle = \cos\left(\theta/2\right)|0\rangle + e^{i \phi}\sin\left(\theta/2\right)|1\rangle$ with a value of $\theta$ which ranges from $0 \ to \ \pi$, while $\phi$ ranges from $0$ to $2\pi$. For~this investigation, we will consider the full range of $\theta$ and $\phi$ values. Here, $| 0\rangle = \begin{pmatrix} 1 \\ 0 \end{pmatrix} $ is the ground state and $| 1 \rangle = \begin{pmatrix} 0 \\ 1 \end{pmatrix} $ is the excited state. In~this paper, energies are given in units of $\epsilon_{2}$, which is then set to 1 in all~calculations.

The total initial Hamiltonian is non-interacting and of the form,
\begin{equation}
\label{eq:systemhamiltonian}
 H_{sys} = \sum_j H_j 
\end{equation}
with $H_j$ given by eq. (\ref{H_j}) and $j=A,B$ and $j=A,B,C$ for two and three qubits, respectively.
Eq. (\ref{eq:systemhamiltonian}) represent the Hamiltonian for the total system at any time, including at the point of measurement, except when NRCGs are applied, inducing interactions between qubits. 
 
To understand the flow of heat through the system, we first define two quantities: the average internal energy and the extractable work. The average internal energy is quantified with the value $ U_j $ and has the form $U_j  = Tr\left[\rho_{j} H_{j}\right]$,

where $\rho_{j}$ is the density matrix of either the component qubits or the total system and $H_{j}$ is the corresponding Hamiltonian, $j = A, \ B, \ C$ for the component qubits or $j = sys$ for the total system. It then follows that the change in energy of (part of) the system from some initial state to some final state is,

\begin{equation}
     \Delta U_j(\rho_{j})  = Tr\left[ \rho^{(f)}_{j} H_{j} \right] - Tr\left[\rho^{(0)}_{j} H_{j}\right], 
\end{equation}
where $\rho^{(0)}_{j}$ and $\rho^{(f)}_{j}$ are the initial and final density matrices respectively. We also define here the convention that a reduction in the total energy of the system is equivalent to a positive \textit{extractable work} $\langle W^e \rangle = -  \Delta U_{sys}$, this definition of extractable work will be used throughout the paper.

We now characterise the flow of heat through the system. Qubits A and C are initialised at a different initial energy to qubit-B, allowing for identification of which components of the working medium are hot and cold, this being important when describing the system as a thermal engine. 
Explicitly,

\begin{itemize}
    \item $U_A^{(0)}=U_C^{(0)}<U_B^{(0)}$ identifies qubit-B as the \textit{hot} component;
    \item $U_A^{(0)}=U_C^{(0)}>U_B^{(0)}$ identifies qubit-B as the \textit{cold} component.
\end{itemize}

Here $U_j^{(0)}$ is the initial energy of the qubits and $j=A, \ B$ and $j=A, \ B, \ C$ for two and three qubits. To identify the direction of the flow of heat, $\langle Q \rangle$ is negative when energy is transferred from a reservoir to a component of the working medium and corresponds to the change in energy of the component qubits, i.e. $\langle Q_j \rangle = \Delta U_j$ where $j = A, \ B$ or $C$. Further, the system has no intrinsic interaction between the qubits outside of those facilitated by NRCGs allowing for convenient resetting of the system back to initial conditions. 

\subsection{Protocol Design and Application}\label{Protocol Application Step}

\begin{figure}[hbt!]
    \centering
    \includegraphics[width=0.7\linewidth]{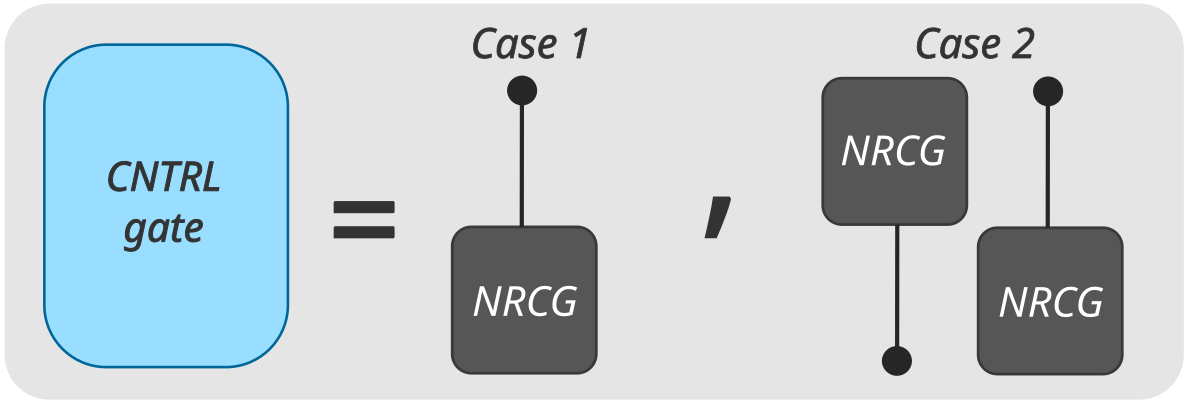}
    \caption{The protocols considered, i.e.,~Case 1 and Case 2, differ by the set of gates within each `controlled gate' block, as~indicated.}
    \label{fig:Circuittwobody}
\end{figure}

The second step of the NRCG driven quantum heat engine is the iterative protocol application shown in Figure~\ref{fig:Circuittwobody}(b). First, we describe  the NRCGs, then the protocols. 

Qubits' correlation which a CNOT gate may induce may lead to a quantum advantage in the operation of a quantum heat engine~\cite{Herrera2023}. The~NRCG is a method of partially applying a CNOT gate; it is a unitary operation given by~\cite{Nikolov},
\begin{equation}\label{NRCGFORM}
\sqrt[N]{CNOT_{A,B}} \ = \
\begin{pmatrix}
1 & 0 & 0 & 0\\
0 & 1 & 0 & 0\\
0 & 0 &  s & p\\
0 & 0 &  p & s
\end{pmatrix}, \;\;\;\;\;\;
\sqrt[N]{CNOT_{B,A}} \ = \
\begin{pmatrix}
1 & 0 & 0 & 0\\
0 & s & 0 & p\\
0 & 0 &  1 & 0\\
0 & p &  0 & s
\end{pmatrix}.
\end{equation}
Here, $ s = \frac{1}{2} + \frac{1}{2}e^{\frac{i\pi}{N}}$, $p = \frac{1}{2} - \frac{1}{2}e^{\frac{i\pi}{N}}$, and $N$ is the number of iterations. The first subscript represents the control qubit and the second subscript, the target qubit. This type of gate may generate entanglement between two qubits, which adds a level of quantumness to the system with the standard form of the CNOT gate recovered when $N = 1$.

A cycle is defined as $N$ iterations of the basic circuit, corresponding to $M=N$ in Figure~\ref{fig:Thermal Machines}(b).
$N$ consecutive applications of an NRCG with the same control and target qubits could be seen as a trotterization of the CNOT gate, aiming at explicitly implementing the gate, in the limit of a large $N$, as an adiabatic dynamic. In~this sense, our protocols give explicit access to intermediate states, e.g.,~allowing for the opportunity to use states with different degrees of entanglement from the end result of the full CNOT~gate. Here $N = 15$ is chosen as it shows good access to intermediate states~\cite{e26110952}.

A schematic of the building blocks of the protocols is shown in Figure \ref{fig:Circuittwobody}: we consider systems of two and three qubits for each protocol and examine two ways of applying interactions. The two-qubit system is recovered with the removal of qubit-C and the second 'controlled gate' block in Figure~\ref{fig:Thermal Machines}(b). 
Explicitly,
\begin{itemize}
\item Case 1: Qubit A is never the target qubit;
\item Case 2: All qubits are either a control or target qubit over the course of one iteration.
\end{itemize}

The system is evolved, according to the quantum circuits in Figure~\ref{fig:Circuittwobody}(b), with the~evolution $\rho^{(f)}_{sys} = \mathcal{U}\rho^{(0)}_{sys}\mathcal{U}^{\dagger}$,

 where $\rho^{(0)}_{sys}$ is the initial density matrix of the total system  and  $\rho^{(f)}_{sys}$ indicates the total density matrix at the end of the protocol. The~unitary $\mathcal{U}$ represents the controlled gates specified in Figure~\ref{fig:Circuittwobody}(b), as~appropriate to each protocol, with~NRCGs of the form in Equation~(\ref{NRCGFORM}).

We note that only for~Case 1 and the two-qubit system, multiples of $N$ iterations of the circuit will be equivalent to the application of multiple standard CNOT gates; however, fractions of $N$ iterations will allow access to intermediate states in the evolution leading to a~CNOT.

These protocols are iteratively applied to the system and can be halted at a chosen number of iterations, which can be optimised for either a maximum or a specific value of $\langle W^e \rangle$. 
Afterwards, the system will then proceed back to the initialisation step. 

When qubit-B is initially the most energetic, we can define qubits A and C as the cold components, and qubit-B as the hot component of the working medium. Identifying a heat engine regime can then be achieved with the fulfilment the following conditions 
\begin{equation}\label{eq:hecond}
    \begin{gathered}
        \Delta U_A \geq 0, \ \Delta U_C \geq 0, \  \Delta U_B < 0, \ \Delta U_{sys} < 0.
    \end{gathered}
\end{equation}
These conditions are specific to qubit-B being the hot component. While the system does not always operate as a heat engine over all protocols and initial conditions, it does so for the regions corresponding to the highest production of extractable work. 

\section{Plan of the Paper and Anticipation of Main~Results}

The application of Nth-root gate operations between qubits  allows access to the system time evolution in a step-wise manner. Consequently, we consider varying the number of iterations and different initialisations for two different gate building blocks, highlighted in Figure~\ref{fig:Circuittwobody}. 

This paper is composed of two main parts. The first part contains an investigation of the systems properties as a heat engine, in some details:

\begin{itemize}
    \item Optimisation of the initialisation parameters $k_b\bar{T}$ for qubits A and C, and $\theta$ and $\phi$ for qubit-B with the aim of maximize production of extractable work;
    \item Comparison of results for qubit-B initialised in a pure state against qubit-B initialised in a Gibbs state for identification of advantages obtained from adding quantumness to the system;
    \item Analysis of the cycles leading to maximum extractable work and of their efficiency.
\end{itemize}

For three out of the four Case and system-size combinations, we find that initial coherence of qubit-B provides an advantage in maximum work production. For all combinations, there is a larger work producing parameter region when compared to a completely classical initial system.  Also, we see that large values of extractable work correspond to large values of efficiency, rendering our system a strong candidate as a quantum heat engine. 

The second part utilises the Pearson Correlation Coefficient (PCC) to quantify the correlation between the extractable work and different correlation measures when the system is initialised to produce maximum work. The correlation measures investigated are:

\begin{itemize}
    \item Mutual information;
    \item Classical correlations;
    \item Quantum discord.
\end{itemize}
 
Our results demonstrate a strong linear correlation between the mutual information and the extractable work, suggesting that correlations play an important role in work production in a quantum heat engine. Of the mutual information components, it is the classical one that correlates better to the extractable work.

\section{Results}

\subsection{Temperature Dependence of Work Production}

\begin{figure}[hbt!]
    \centering
    \includegraphics[width=15.5cm]{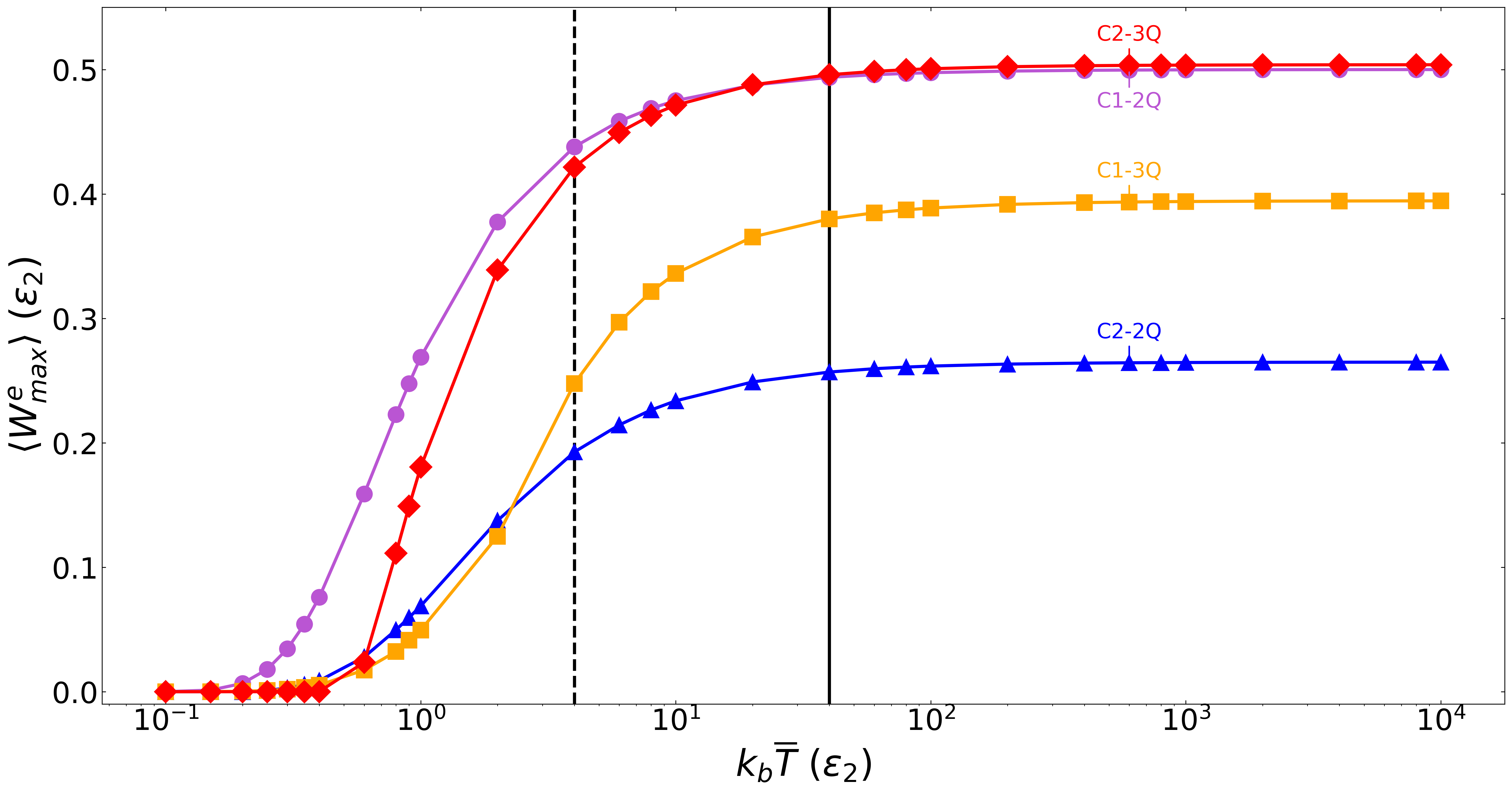}
    \caption{Results demonstrate the $k_b\bar{T}$ dependence in units of $\epsilon_{2}$ (x-axis) of the maximum extractable work $\langle W^e_{max} \rangle$ in units of $\epsilon_{2}$ (y-axis), where $\langle W^e_{max} \rangle$ is recorded over every $\theta$ and $\phi$ combination per value of $k_b\bar{T}$. Black vertical line: $k_b\bar{T} = 40\epsilon_{2}$. Dashed black vertical line: $k_b\bar{T} = 4\epsilon_{2}$. Purple: Case 1 two-qubit; Blue: Case 2 two-qubit; Orange: Case 1 three-qubit; Red: Case 2 three-qubit, as labelled.}
    \label{fig:Temp Dependence}
\end{figure}

Here, we begin with optimising the system described in Section \ref{theoreticaloutline} to identify the temperatures at which to initialise qubits A and C for maximum extractable work $\langle W^e_{max} \rangle$. This is defined as the peak value of extractable work across all protocols' applications and system initialisations. Figure \ref{fig:Temp Dependence} shows the temperature dependence of $\langle W^e_{max} \rangle$ from all protocols and system sizes, where protocols are applied up to 150 iterations, $\theta$ is scanned from $0 \le \theta \le \pi$, $\phi$ is scanned from $0 \le \phi \le 2\pi$, a comprehensive range of $k_b\bar{T}$ is sampled, and $\langle W^e_{max} \rangle$ measured for each combination.

\begin{table*}[hbt!]
\centering
\begin{tabular}{c c c| c| c}
        &                            & \multicolumn{2}{c}{$\langle W^e_{max} \rangle$ at $k_b\bar{T} = 10^4\epsilon_{2}$}&   \\ \cline{3-4} 
        & \multicolumn{1}{c|}{}       &\multicolumn{1}{c|}{\ \ \  Two-Qubit \ \ \ }          & \multicolumn{1}{c|}{\ \ \  Three-Qubit \ \ \ }  & \\ \cline{3-4} 
        & \multicolumn{1}{c|}{Case 1} & $0.50\epsilon_{2}$ & $0.39\epsilon_{2}$&  \\ \cline{3-4} 
        & \multicolumn{1}{c|}{Case 2} & $0.27\epsilon_{2}$ & $0.50\epsilon_{2}$&   \\  \cline{3-4} 
\end{tabular}
\caption{\label{table:Temp Convergence}Maximum extractable work as converged at large values of $k_b\bar{T}$, corresponding to Figure \ref{fig:Temp Dependence}.}
\end{table*}

Higher initial temperatures lead to larger values of $\langle W^e_{max} \rangle$, until saturation occurs for $k_b\bar{T} \gtrsim 40\epsilon_{2}$. This is intuitive, the system is initialised with a higher initial energy leaving more scope for a larger reduction in energy and a larger amount of work to be produced. Low temperatures show small to no work production until $k_b\bar{T} \approx 0.4\epsilon_{2}$ for all but  Case 1 two-qubit system, which produces some amount of work already at $k_b\bar{T} \approx 0.1\epsilon_{2}$.
Work production then rapidly increases up to $k_b\bar{T} \approx 4\epsilon_{2}$, to saturate at higher temperatures to the values of peak work shown in Table \ref{table:Temp Convergence} for each protocol and system size. This convergence is a result of the thermally initialised qubits approaching maximally mixed states. Because of this, the temperature at which we see the largest $\langle W^e_{max} \rangle$ is the same for all Cases and system sizes. Figure \ref{fig:Temp Dependence} already demonstrates that the maximum extractable work is not merely connected to the system size and available initial energy, as, depending on the protocol, the two- or three-qubit system may achieve a larger $\langle W^e_{max} \rangle$. As this work will focus mainly on maximising work production, in the following we choose to initialise the system with $k_b\bar{T} = 40\epsilon_{2}$ which is shown in Figure \ref{fig:Temp Dependence} by the black vertical line. For completeness an investigation into $k_b\bar{T} = 4\epsilon_{2}$, signified by the black dashed vertical line, was also performed, and showed similar behaviour to that of $k_b\bar{T} = 40\epsilon_{2}$. 

\subsection{Maximum Extractable Work Dependence on Initial Coherences}\label{Maxworkscansthermalcomparrison}

Following from finding optimal initial temperature for $k_b\bar{T}$ we now investigate the $\theta$ and $\phi$ dependence of $\langle W^e_{max} \rangle$. A range of initialisation values of $\theta$ and $\phi$ are scanned from $0 \leq \theta \leq \pi$ and $0 \leq \theta \leq 2\pi$ respectively with each combination having its respective protocol applied for up to 150 iterations.

\begin{figure}[hbt!]
    \centering
    \includegraphics[width=15.5cm]{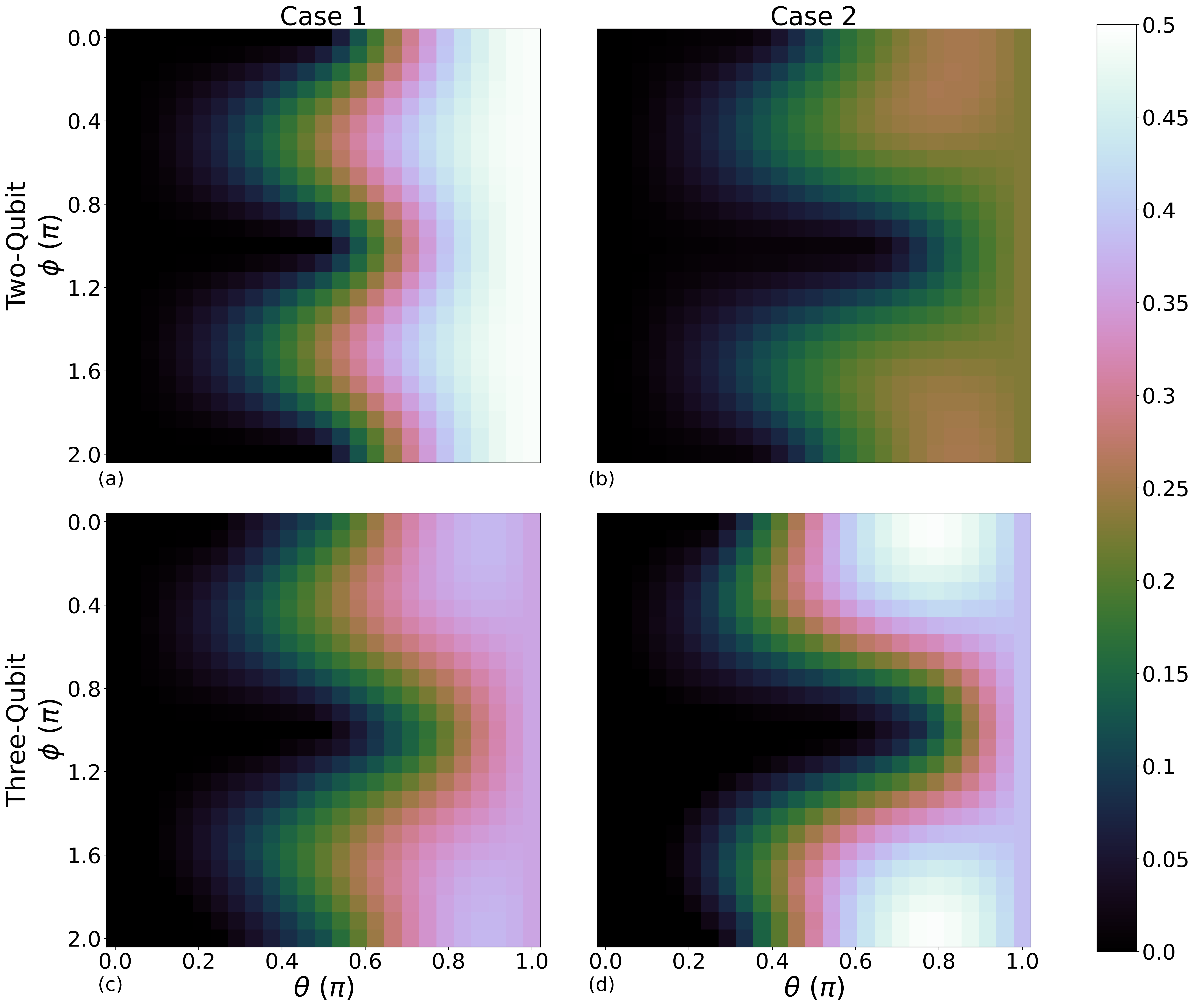}
    \caption{Maximum extractable work $\langle W^e_{max} \rangle$ in units of $\epsilon_{2}$ for each $\{\theta, \phi \}$ combination and $0 \leq \theta \leq \pi$ (x-axis) and $0 \leq \phi \leq 2\pi$ (y-axis); First row: Two-qubit systems where columns left to right are cases 1 and 2 respectively. Brighter shades correspond to a greater maximum value of extractable work $\langle W^e \rangle$. Each combination of initial $\theta$ and $\phi$ is run for 150 iterations with $\langle W^e_{max} \rangle$ recorded at each iteration. Parameters are $\epsilon_{1} = 0\epsilon_{2}$ and $k_b\bar{T} = 40\epsilon_{2}$. Second row: The same parameters as the first row but for the three-qubit systems.}    
    \label{fig:Maxworkscansphi}
\end{figure}

\begin{table*}[hbt!]  
\centering
    \begin{tabular}{ c c c|c|c| c c|c|c| c}
             &                                 & \multicolumn{3}{c}{Two-Qubit} & & \multicolumn{3}{c}{Three-Qubit} & \\\cline{3-5} \cline{7-9}
             &\multicolumn{1}{c|}{}            & \multicolumn{1}{c|}{\ \ $\theta$ \ \ } & \multicolumn{1}{c|}{\ \ $\phi$ \  } & \multicolumn{1}{c|}{\ \ $\langle W^e_{max} \rangle$ \ \ } & \multicolumn{1}{c|}{}& \multicolumn{1}{c|}{\ \ $\theta$ \ \ } & \multicolumn{1}{c|}{\ \ $\phi$ \ \ } & \multicolumn{1}{c|}{\ \ $\langle W^e_{max} \rangle$ \ \ }\\ \cline{3-5} \cline{7-9}
             &\multicolumn{1}{c|}{Case 1}   & $\pi$ & any & $0.49\epsilon_{2}$ & \multicolumn{1}{c|}{}& $0.88\pi$ & $0.24\pi$ & $0.38\epsilon_{2}$ \\ \cline{3-5} \cline{7-9}
             &\multicolumn{1}{c|}{Case 2} & $0.83\pi$ & $0.32\pi$ & $0.26\epsilon_{2}$ & \multicolumn{1}{c|}{}& $0.79\pi$ & $0.08\pi$ & $0.50\epsilon_{2}$ \\ \cline{3-5} \cline{7-9}
    \end{tabular}
    \caption{\label{table:maxwork}Maximum work values from Figure \ref{fig:Maxworkscansphi} and their corresponding initialisation parameters for Case 1 and 2. Left: Two-qubit; Right: Three-qubit.}
\end{table*}

The Case 1 two-qubit and Case 2 three-qubit systems, Figure \ref{fig:Maxworkscansphi}(a) and Figure \ref{fig:Maxworkscansphi}(d) respectively, demonstrate the largest $\langle W^e_{max} \rangle$, being the most favourable system combinations for use as an engine. The shape of the work producing regions provides some insight into the initialisation parameters that lead to large work production for our protocol. 

When $\theta = 0.5\pi$ qubit-B has the same initial energy as qubits-A and -C: as $\theta$ increases qubit-B becomes the initially hot component of the working medium. Maximum work production always resides in the region where $\theta > \pi/2$, though work can still be produced when $\theta < \pi/2$. Table \ref{table:maxwork} shows the initialisation parameters for peak work production for each Case and both two- and three-qubit systems. Apart from the Case 1 two-qubit system, we see that there is also a $\phi$ dependence on $\langle W^e_{max} \rangle$. However, even for the Case 1 two-qubit system there are values of $\phi$ for which the work producing region extends to $\theta < 0.5\pi$. For all cases, the topography of these regions are almost mirrored around $\phi = \pi$ though the largest $\langle W^e_{max} \rangle$ by a small margin is found when $\phi < \pi$. 
Much like for large values of $k_b\bar{T}$ discussed in the previous section, intuitively, larger initial values of $\theta$ would produce the most work. Increasing $\theta$ increases qubit-B's initial energy from $\epsilon_{1}=0$ to $\epsilon_{2}$, getting the system close to its maximum possible initial energy.
Indeed, large initialisation values of $\theta$ are beneficial but in three out of the four cases the largest values of $\langle W^e_{max}\rangle$ are found at $\theta < \pi$, highlighting the positive role of initial coherences. 

\subsection{Comparison to initialization with no coherences}\label{Comparison to initialization with no coherences}

We now compare the $\langle W^e_{max} \rangle$ achieved when qubit-B is initialised with coherence, against qubit-B initialised in a thermal state without initial coherence. Effective negative temperatures are included when considering this comparison to achieve equivalent initial ground and excited state populations for qubit-B. The corresponding 'thermal' plots to Figure~\ref{fig:Maxworkscansphi} are included in Appendix~\ref{appendB} where, in Figure~\ref{fig:Completely thermal maximum work} each point in the $\{\theta,\phi\}$ parameter region has the same initial energy as in Figure~\ref{fig:Maxworkscansphi} and qubit-B is initialised in the Gibbs state corresponding to that energy. The thermal initialisation produces maximum work when qubit-B is initialised entirely in its excited state ($\theta = \pi$) for all case and system size combinations. This is in contrast to Figure~\ref{fig:Maxworkscansphi}, where for three out of the four case and system size combinations $\langle W^e_{max} \rangle$ is achieved when $\theta < \pi$, and qubit-B has initial coherence. 

Figure \ref{fig:Maxworkscansphi} shows a beneficial $\phi$ dependence on the work-producing parameter region, as for $\phi \approx 0.5\pi$ and $\phi \approx 1.5\pi$ the work producing parameter regions extend to lower values of $\theta$ than seen in the completely thermal initialisation. This is true for all cases and system sizes. 

\begin{table}[hbt!] 
\centering
    \begin{tabular}{c  c c|c|c c|c|  c}
             &  & \multicolumn{2}{c}{$\langle W^e_{max} \rangle$} & & \multicolumn{2}{c}{Relative change in $\langle W^e_{max} \rangle$} & \\\cline{3-4} \cline{6-7}
             &\multicolumn{1}{c|}{}            & \multicolumn{1}{c|}{\ \ \  Two-Qubit \ \ \ }          & \multicolumn{1}{c|}{\ \ \  Three-Qubit \ \ \ }         & \multicolumn{1}{c|}{}& \multicolumn{1}{c|}{\ \ \  Two-Qubit \ \ \ } & \multicolumn{1}{c|}{\ \ \  Three-Qubit \ \ \ } & \\ \cline{3-4} \cline{6-7}
             &\multicolumn{1}{c|}{Case 1}      & $0.49\epsilon_{2}$ & $0.36\epsilon_{2}$  & \multicolumn{1}{c|}{}& $0\%$     & $5.56\%$    & \\ \cline{3-4} \cline{6-7}
             &\multicolumn{1}{c|}{Case 2}      & $0.23\epsilon_{2}$ & $0.39\epsilon_{2}$  & \multicolumn{1}{c|}{}& $12.04\%$ & $28.20\%$   & \\\cline{3-4} \cline{6-7}
    \end{tabular}
    \caption{\label{table:Completely thermal maximum work}Left: maximum work values for Case 1 and 2 for the two- and three-qubit system with a complete thermal initialisation; Right: percentage change in $\langle W^e_{max} \rangle$ by introducing initial coherence in qubit-B with a pure state initialisation.}
\end{table}

The $\langle W^e_{max}\rangle$ for the completely thermal case and its increase  between the thermal and pure-state initialisation of qubit-B is shown in Table \ref{table:Completely thermal maximum work}, left and right tables respectively. Here, we observe an improvement in $\langle W^e_{max} \rangle$, when initialising qubit-B in a pure state, for all but the Case 1 two-qubit system, for which 
the maximum work producing initialisation parameters with coherences correspond to $\theta = \pi$. This renders its initial state equivalent to the maximum work producing initial state for the completely thermal system, hence no increase to the value of $\langle W^e_{max} \rangle$. The greatest improvement in maximum work production for the Case 2 three-qubit system, yielding an increase of $25.64\%$. 

We also checked performance against the use of standard -- instead of NRCGs -- CNOT gates in the protocols: for some parameter regions, NRCGs produces larger peak extractable work than standard CNOT gates, on top of the ability to access intermediate work producing states not readily accessible by full CNOT gates. Some examples are in Appendix \ref{appendC}.

\subsection{Energy Dynamics when Operating within a Maximum Work Producing Regime}

\begin{figure}[hbt!]
    \centering
    \includegraphics[width=15.5cm]{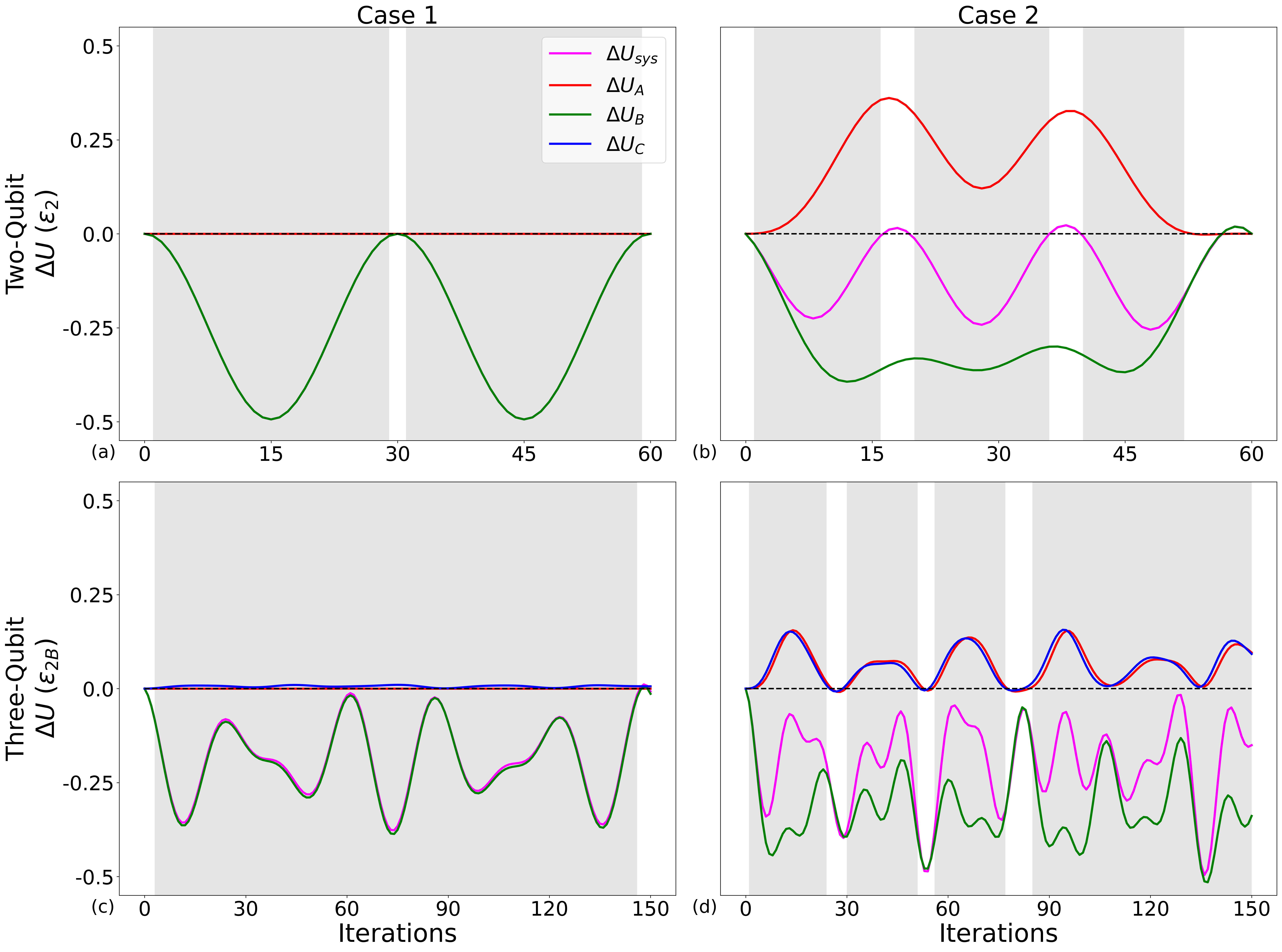}
    \caption{Change in average internal energy $\Delta U$ for the total system and component qubits in units of $\epsilon_{2}$ (y-axis) for $0 \leq $ iterations $ \leq 60$ for two-qubit systems (first row) and $0 \leq $ iterations $ \leq 150$ for three-qubit systems (second row) (x-axis), where columns from left to right are Case 1 and 2 respectively. Corresponding initial $\theta$ and $\phi$ for each panel is shown in table \ref{table:maxwork}, with $\phi=0.17\pi$ for Case 1 two-qubits. Grey shaded regions identify heat engine operation. Same parameters as in Figure \ref{fig:Maxworkscansphi}. Red line: $\Delta U_A$; Green line: $\Delta U_B$; Blue line: $\Delta U_C$; Magenta line: $\Delta U_{sys}$.}    
    \label{fig:Maxworkscans}
\end{figure}

In Figure~\ref{fig:Maxworkscans} we plot the detailed dynamics of the whole system energy variation $\Delta U_{sys}$ and of its components  $\Delta U_A$,  $\Delta U_C$,   $\Delta U_B$ for initializations corresponding to the maximum work values shown in Table \ref{table:maxwork}. For Case 1 two-qubits, we choose $\phi=0.17\pi$, for which the work producing region extends to $\theta<\pi/2$.
The two-qubit systems are run over 60 iterations as any more and we observe repeated energy dynamics while, the three-qubit systems are run over 150 iterations. In all panels qubit-B is initially the most energetic, with its excited state having a higher initial population than qubits A and C.  The grey shaded regions in Figure~\ref{fig:Maxworkscans} show when the system is operating in a heat engine regime according to the conditions in Eq.~(\ref{eq:hecond}), and will be used for the same purpose in all relevant figures throughout the rest of this work. The Case 2 three-qubit system shows that producing work, i.e. $\Delta U_{sys}<0$, does not necessarily indicate that the system is operating as a heat engine, with the second largest peak work falling outside the shaded area. However, most of the time for all Cases and system sizes we generally operate in the heat engine regime. We also see, in general, that an increase in energy of qubits A and C corresponds to a reduction in energy of qubit-B. The exception is the Case 1 two-qubit system. As qubit-A is only ever a control qubit, its initial (local) state is preserved through the application of the protocol with the system demonstrating no flow of heat from the hot component of the working medium to the cold, though it still produces work.

$\Delta U_A$ and $\Delta U_C$ follow very similar dynamics and, for Case 1, are always almost zero. Their variations for Case 2 are due to the presence of the extra gates. This is reflected in $\Delta U_{sys}$, which, for Case 2, does not always track $\Delta U_B$ over the course of repeated iterations. Qubit-B remains the main contributor to the energy dynamics of the total system.  

\subsection{Efficiency}

\begin{figure}[hbt!]
    \centering
    \includegraphics[width=15.5cm]{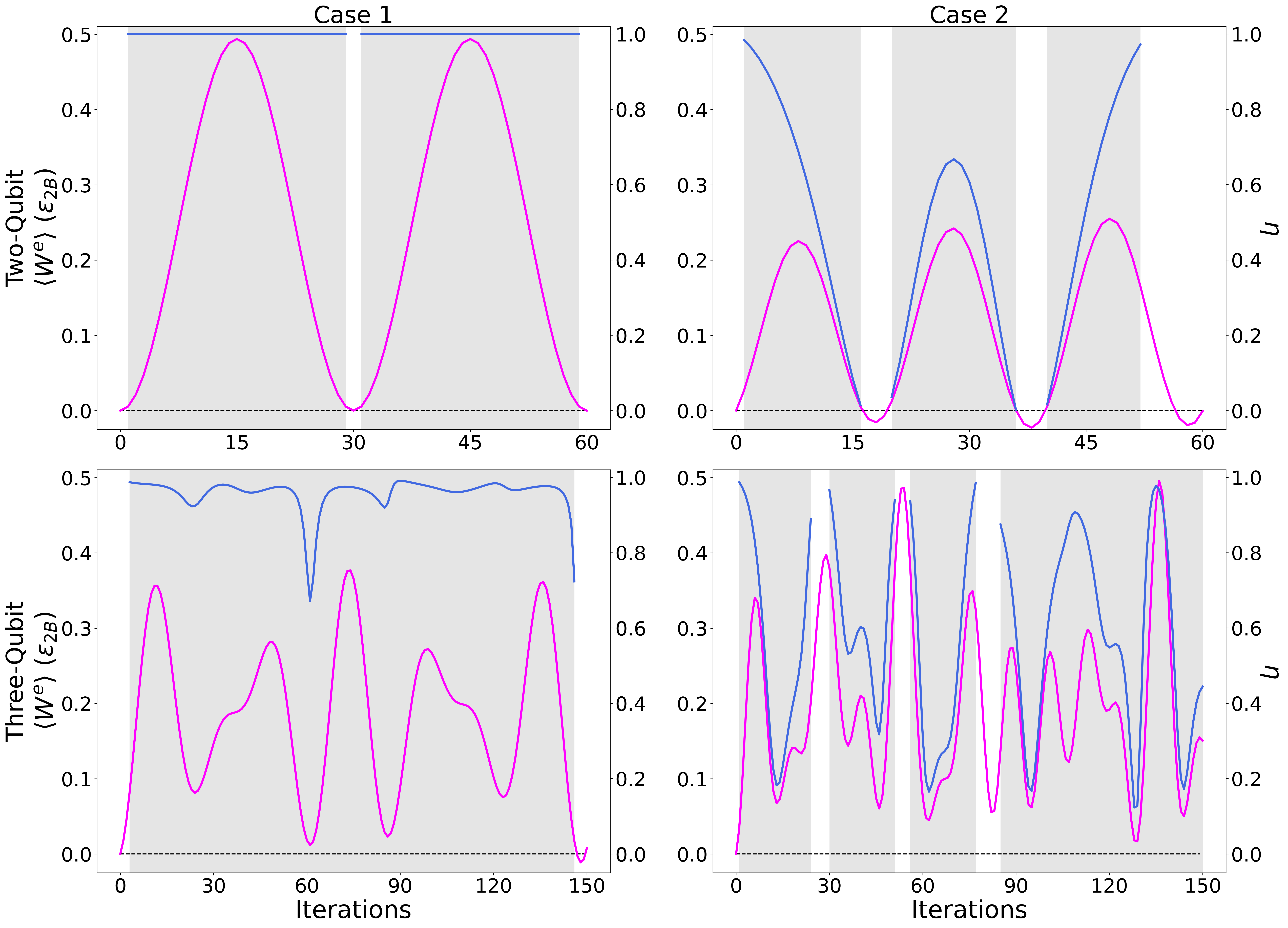}
    \caption{Comparing the extractable work $\langle W^e\rangle=-\Delta U_{sys}$ (magenta) in units of $\epsilon_{2}$ (left y-axis) and the efficiency  $\eta$ (blue, right y-axis), in the case of the two-qubit systems (first row) for $0 \leq $ iterations $ \leq 60$ and the three-qubit systems (second row) for $0 \leq $ iterations $ \leq 150$, where columns left to right are Case 1 and 2 respectively. Same parameters as in Figure \ref{fig:Maxworkscans}.  Magenta: $\langle W^e\rangle$; Blue: $\eta$.}    
    \label{fig:efficiencymaxwork}
\end{figure}

Efficiency is calculated as $\eta = - \langle W^e \rangle / Q_{H}$ where $\langle W^e \rangle$ is the extractable work and $Q_{H} =  \Delta U_{H} $ the energy difference of the hot component of the working medium. Comparison of the work and corresponding efficiency is shown in Figure \ref{fig:efficiencymaxwork} with initialisation parameters corresponding to the maximum work scans in Figure \ref{fig:Maxworkscans}. All Cases and system sizes are considered. Efficiency is only shown when the system satisfies conditions (Equation~\ref{eq:hecond}) that indicate its operation as a heat engine (grey shaded areas). Table \ref{table:efficiency} shows the efficiency corresponding to the $\langle W^e_{max} \rangle$ in figure \ref{fig:efficiencymaxwork} for each Case and system size.

\begin{table}[hbt!]
\centering
    \begin{tabular}{c c c|c| c}
        &                             & \multicolumn{2}{c}{$\eta$ at $\langle W^e_{max}\rangle$}& \\ \cline{3-4} 
        &    \multicolumn{1}{c|}{}    & \multicolumn{1}{c|}{\ \ \  Two-Qubit \ \ \ } & \multicolumn{1}{c|}{\ \ \  Three-Qubit \ \ \ } &\\ \cline{3-4} 
        & \multicolumn{1}{c|}{Case 1} & $1.00$  & $0.84$& \\ \cline{3-4} 
        & \multicolumn{1}{c|}{Case 2} & $0.97$ & $0.93$& \\ \cline{3-4} 
    \end{tabular}
    \caption{\label{table:efficiency}Efficiency corresponding to the maximum extractable work, corresponding to Figure \ref{fig:efficiencymaxwork}.}
\end{table}

The Case 1 two-qubit (Figure~\ref{fig:efficiencymaxwork}(a)) differs from of all others as $\Delta U_A = 0$ during the protocol. This results in an efficiency of 1 being achieved as $100\%$ of $\langle Q_H \rangle$ is imparted onto qubit-B and is extractable as work. 
For all other cases, the efficiency varies as the protocols are applied. Both Case 2 systems show a large variation in the values of efficiency, reflecting the larger energy variation of qubits A and C.  Notably, peak values of $\langle W^e \rangle$ are marked by peak values of efficiency and/or large efficiency.
The more complex efficiency behaviour for Case 2 is to be expected due to the added complexity of the protocol. Case 2 has all qubits, at some point, being the target of an NRCG whereas Case 1 has one qubit that is never a target and retains its initial energy. This, coupled with doubling the number of gate applications per protocol for the same system size gives rise to this added complexity. 

Comparing with results in Figure \ref{fig:Maxworkscans} the high efficiency peaks occur when $\Delta U_A$ and $\Delta U_C$ approach zero and $\Delta U_B$ is the dominant component of the work. The Case 1 three-qubit system (Figure~\ref{fig:efficiencymaxwork}(c)) maintains a very high value of efficiency while in the heat engine regime. The efficiency never drops below 0.62 while operating mostly between 0.92 and 0.98, displaying the second most stable efficiency after the Case 1 two-qubit system. Positively, halting the protocol at any of the peak work-producing iterations for this Case results in favourable efficiency. Having regions the engine could be halted that both have large work and efficiency is a positive indicator of the viability of this system's use as a quantum heat engine.

\section{Relationship between System's Correlations and Extractable Work}\label{PCCQDMICCMAX}

We utilise the Pearson Correlation Coefficient (PCC) to quantify the strength of the relationship between a number of correlation measures and extractable work. The PCC is a measure of the linear relationship between two data sets, a value approaching 1 signifies a strong linear correlation whereas a value approaching -1 signifies a strong anti-linear one. The PCC has the form
\begin{equation}
    r_{x,y} = \frac{\sum^{n}_{i = 1}\left(x_i - \bar{x}\right)\left(y_i - \bar{y}\right)}{\sqrt{\sum^{n}_{i = 1}\left(x_i - \bar{x}\right)^2}\sqrt{\sum^{n}_{i = 1}\left(y_i - \bar{y}\right)^2}},
\end{equation}
where $x_i$ and $y_i$ are dataset values, $\bar{x}$ and $\bar{y}$ are the corresponding mean values, and $n$ is the length of the data sets $x$ and $y$. We are interested in $r$ values where $r < -0.5$ and $r > 0.5$, corresponding to a level of significant correlation.

We now introduce and define the three quantities we will investigate for the remainder of the paper. The mutual information, total classical correlations, and the quantum discord. The mutual information \cite{PhysRevA.71.062307} captures all information shared between the components of a chosen bi-partition encapsulating both classical and quantum correlations and is given by

\begin{equation}
    I(A:B) = S(\rho_A) + S(\rho_B) - S(\rho_{AB}).
    \label{eq:mutualinformation}
\end{equation}
Here $S(\rho_j) = -Tr[\rho_j \ln{\rho_j}]$ \cite{Nielsen_Chuang_2010,Peres1995-PEREQT} is the von Neumann entropy, $A,~B$ signifies the components of the chosen bi-partition, and $\rho_A,~\rho_B$, the corresponding reduced density matrices traced from the system density matrix $\rho_{AB}$. The quantum discord~\cite{PhysRevLett.88.017901} is a measure of total quantum correlations, 
\begin{equation}
    D(A:B)_{\{\Pi_j^B \}} = \min_{\Pi^B_j}[ S(\rho_B) - S(\rho_{AB}) + S(\rho_A|\{ \Pi ^B_j\})],    
    \label{eq:discord}    
\end{equation}
where $ \Pi ^B_j$ are the projective measurements on $\rho_B$ and $S(\rho_A|\{\Pi_j^B \}) = \sum_{j = u, v}  p_j S(\rho_{A| \Pi^B_j})$ is the conditional entropy. Here, the post measurement state of $A$ that corresponds to the outcome of the measurement on $B$ has the form 

\begin{equation}
    \begin{gathered}
        \rho_{A| \Pi^B_j} = ( \Pi^B_j) \rho_{AB} ( \Pi^B_j) / p_j, \ \ p_j = tr[\Pi^B_j 
        \rho_{AB}]. 
    \end{gathered}
\end{equation}

The complete orthonormal measurement basis $\{ \ket{u}, \ket{v} \}$ is $\{ \ket{u} = \cos{(\theta/2)} \ket{0} + \sin{(\theta/2)}e^{i\phi}\ket{1}, \ket{v} = \sin{(\theta/2)}e^{-i\phi}\ket{0} - \cos{(\theta/2)} \ket{1}\}$ and is used to create projectors for the measurement $\Pi^B_u = \mathbb{I}^A \otimes \ket{u}\bra{u}$ and $\Pi^B_v = \mathbb{I}^A \otimes \ket{v}\bra{v}$, where $\mathbb{I}^A$ is the identity matrix with dimensionality corresponding to that of subsystem A. To satisfy the minimisation of the discord, measurement is performed for all combinations of $0 \leq \theta \leq \pi$ and $0 \leq \phi \leq 2\pi$. A full derivation of the discord and its adaption to the three-qubit case is provided in Appendix \ref{appendD}. 

Completely classical correlations are also considered. As the discord encapsulates all the quantum correlations in the mutual information then it follows that the remaining correlations are completely classical and given by

\begin{equation}
\begin{gathered}
    CC(A:B) = I(A:B) - D(A:B)_{\{\Pi_j^B \}}.
    \label{eq:classicalcorrelation}
\end{gathered}
\end{equation}

We note that $CC(A:B)$ and $D(A:B)_{\{\Pi_j^B \}}$  are not symmetric under the exchange $A\leftrightarrow B$.

\subsection{Relationship between Correlations and Extractable Work}

\subsubsection{Quantum Mutual Information} 
\begin{figure}[hbt!]
    \centering
    \includegraphics[width=15.5cm]{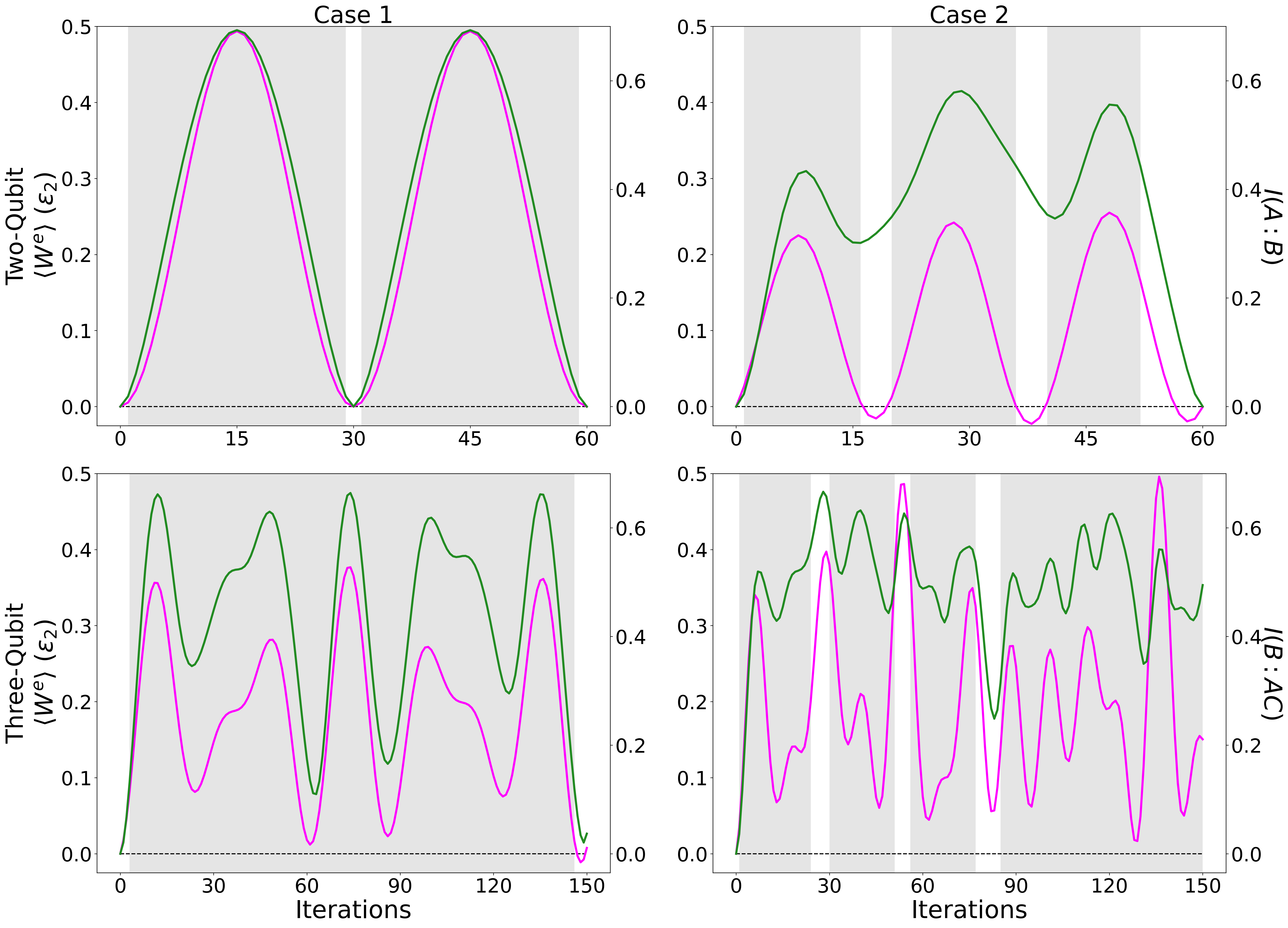}
    \caption{Comparing the extractable work $\langle W^e\rangle$ (magenta) in units of $\epsilon_{2}$ (left y-axis) and the mutual information (green) (right y-axis), in the case of the two-qubit systems (first row) for $0 \leq $ iterations $ \leq 60$ and the three-qubit systems (second row) for $0 \leq $ iterations $ \leq 150$, where columns left to right are Case 1 and 2 respectively. Same parameters as in Figure \ref{fig:Maxworkscans}. Magenta: $\langle W^e\rangle$; Green: Mutual information.}
    \label{fig:MIMaxWorkSlices}
\end{figure}

As shown in Fig.~\ref{fig:Maxworkscansphi} and described in Sections \ref{Maxworkscansthermalcomparrison} and \ref{Comparison to initialization with no coherences},  initial coherence and a finite $\phi$ was beneficial towards peak work production. We now aim to characterise how correlations between bi-partitions of the  working medium, and generated through protocols' applications, are related to the systems performance. Initial $\theta$ and $\phi$ for Figure \ref{fig:MIMaxWorkSlices} are the same as in Figure \ref{fig:Maxworkscans}. The selected bi-partition for the two-qubit system is straight forward to choose, with only a single option $(A:B)$, whereas for the three-qubit system we have three possible bi-partitions. We choose $(AC:B)$ due to qubit-B interacting with both qubits A and C directly and to its generally dominant contribution to $\langle W^e \rangle$. 

\begin{table}[hbt!]       
\centering
    \begin{tabular}{cc c|c|c|c c|c|c|c}
         &                                 & \multicolumn{3}{c}{$\langle W^e \rangle$ versus Mutual Information} &                      &\multicolumn{3}{c}{$\Delta U_B$ versus Mutual Information}&\\ \cline{3-5} \cline{7-9}
         &\multicolumn{1}{c|}{}            &  \multicolumn{1}{c|}{\ \ \  Bi-partition \ \ \ }                & \multicolumn{1}{c|}{\ \ \ \ \  Case 1 \ \ \ \ \ } & \multicolumn{1}{c|}{\ \ \ \ \  Case 2 \ \ \ \ \ }    &    \multicolumn{1}{c|}{}    & \multicolumn{1}{c|}{\ \ \  Bi-partition \ \ \ }          & \multicolumn{1}{c|}{\ \ \ \ \  Case 1 \ \ \ \ \ } & \multicolumn{1}{c|}{\ \ \ \ \  Case 2 \ \ \ \ \ }&\\  \cline{3-5} \cline{7-9}
         &\multicolumn{1}{c|}{Two-Qubit}   & $(A:B)$  & 0.99   & 0.67               & \multicolumn{1}{c|}{}& $(A:B)$  & -0.99  & -0.77 & \\ \cline{3-5} \cline{7-9}
         &\multicolumn{1}{c|}{Three-Qubit} & $(AC:B)$ & 0.92   & 0.50               & \multicolumn{1}{c|}{}& $(AC:B)$ & -0.92  & -0.59 & \\ \cline{3-5} \cline{7-9}
    \end{tabular}
    \caption{\label{tab:MIPrMaxWork}Left: PCC for $\langle W^e \rangle$ versus Mutual Information corresponding to Figure \ref{fig:MIMaxWorkSlices}. Right: PCC for $\Delta U_{B}$ versus Mutual Information.}
\end{table}

PCCs for all Cases and system sizes are shown in Table \ref{tab:MIPrMaxWork}. A strong linear correlation between the mutual information and the extractable work is seen in Figure \ref{fig:MIMaxWorkSlices} for the Case 1 two- and three-qubit system and the Case 2 two-qubit system. The weakest  relationship emerges from the Case 2 three-qubit system having a PCC of 0.50. However, even with this lower  value of PCC we still find a strong correspondence between the peaks of extractable work and those of the mutual information. For Case 1 (left column) the extractable work is seen to be directly proportional to the mutual information with larger values of mutual information indicating an equally larger value of work. Case 2 is a more complicated picture. While, much like in Case 1, the peaks in mutual information and extractable work are in agreement, the two quantities are strongly correlated but not as directly proportional. This suggests that the additional inverted gates for Case 2 protocol complicate the work mutual information relationship. Even here though, data suggest that the building of correlations is beneficial to work production.  We find that the mutual information and extractable work for $k_b\bar{T} = 4\epsilon_{2}$ demonstrates  similar linear correlations to the $k_b\bar{T} = 40\epsilon_{2}$ case, reinforcing the validity of these conclusions. 

Strong linear correlation is seen between $\Delta U_B$ and the mutual information (Table \ref{tab:MIPrMaxWork}, right) for each protocol and system size. As the mutual information tracks the work then this is to be expected, with qubit-B generally being largest contributor to the work for all cases and system sizes of Figure \ref{fig:Maxworkscans}. This is especially true for the Case 1 two-qubit system as the work is only governed by $\Delta U_B$. Apart from the sign, there is no different in PCC for Case 1 when comparing between work and $\Delta U_{B}$; instead, Case 2 for both two- and three-qubit systems shows an increase in PCC of $10\%$ and $18\%$, respectively, indicating that the strongest correlation is between the mutual information and $\Delta U_{B}$, which itself has the largest impact on the work produced. 

As the mutual information is a measure of total correlations between sections of a bipartition, quantifying the individual classical and quantum contributions to the systems behaviour in terms of work production may provide a more detailed insight. First, we examine the classical correlations followed by the quantum discord.     

\subsubsection{Classical Correlations and Quantum Discord}

\begin{figure}[hbt!]
    \centering
    \includegraphics[width=15.5cm]{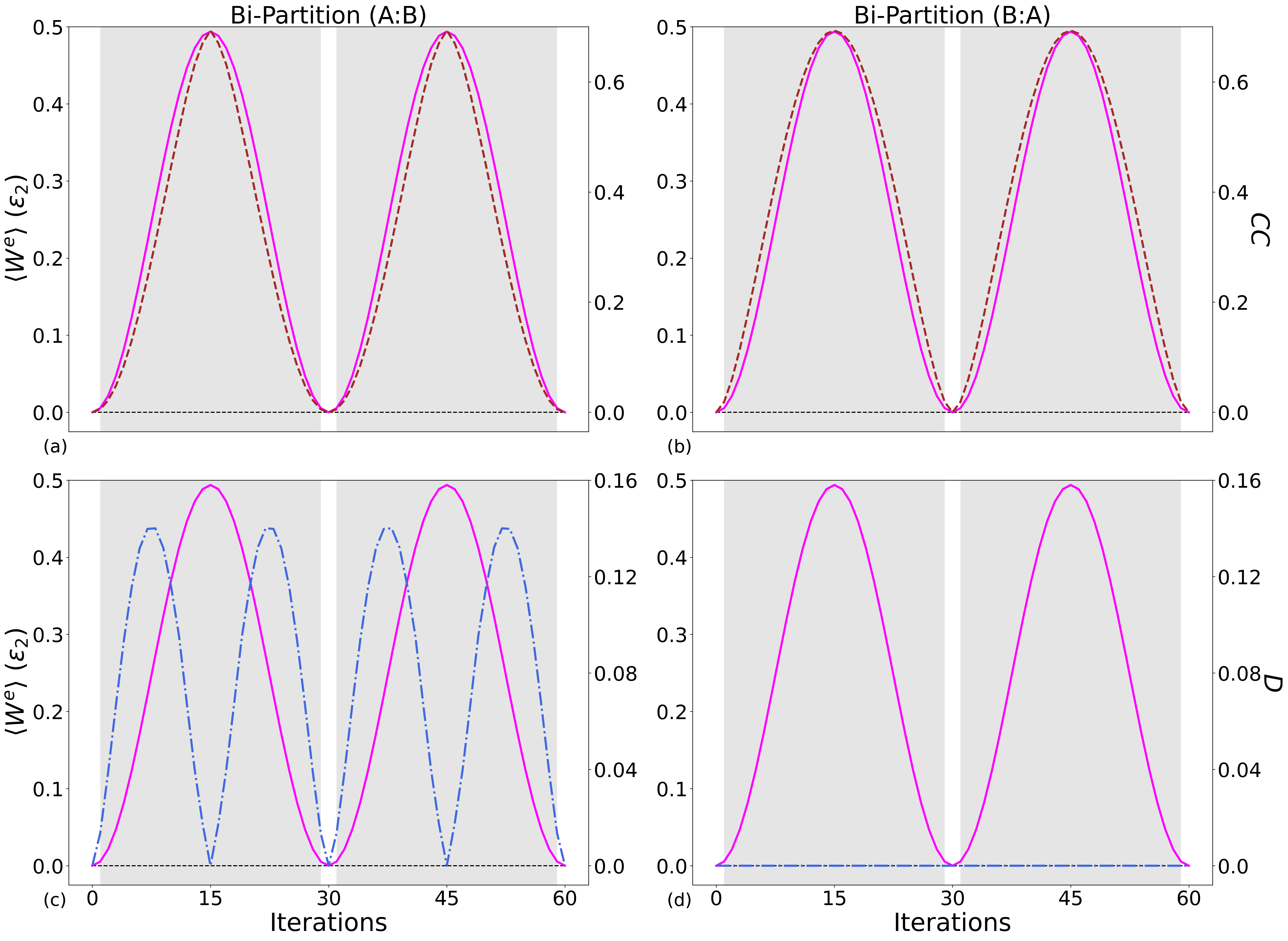}
    \caption{Comparing, for the Case 1 two-qubit system, the extractable work $\langle W^e\rangle$ (magenta lines) in units of $\epsilon_{2}$ (left y-axis), the classical correlations (brown dashed lines, top row right y-axis), and the quantum discord (blue dash dot lines, bottom row right y-axis), for $0 \leq $ iterations $ \leq 60$, where columns left to right are the bi-partitions when measuring on $\rho_B$ and $\rho_A$ respectively.  Same parameters as in Figure \ref{fig:Maxworkscans}. }    
    \label{fig:C12BCCandD}
\end{figure}

\begin{table*}[hbt!]
\centering
    \begin{tabular}{cc c|c|c|c c|c|c|c}
         &                                 & \multicolumn{3}{c}{$\langle W^e \rangle$ versus Classical Correlations} &                      &\multicolumn{3}{c}{$\Delta U_B$ versus Classical Correlations}&\\ \cline{3-5} \cline{7-9}
         &\multicolumn{1}{c|}{}            & \multicolumn{1}{c|}{\ \ \  Bi-partition \ \ \ }                 & \multicolumn{1}{c|}{\ \ \ \ \  Case 1 \ \ \ \ \ } & \multicolumn{1}{c|}{\ \ \ \ \  Case 2 \ \ \ \ \ }             & \multicolumn{1}{c|}{}& \multicolumn{1}{c|}{\ \ \  Bi-partition \ \ \ }          & \multicolumn{1}{c|}{\ \ \ \ \  Case 1 \ \ \ \ \ } & \multicolumn{1}{c|}{\ \ \ \ \  Case 2 \ \ \ \ \ }&\\  \cline{3-5} \cline{7-9}
         &\multicolumn{1}{c|}{Two-Qubit}   & $(A:B)$  & 0.99   & 0.67               & \multicolumn{1}{c|}{}& $(A:B)$  & -0.99  & -0.64 & \\ \cline{3-5} \cline{7-9}
         &\multicolumn{1}{c|}{Two-Qubit} & $(B:A)$ & 0.99   & 0.73               & \multicolumn{1}{c|}{}& $(B:A)$ & -0.99  & -0.38 & \\ \cline{3-5} \cline{7-9}
    \end{tabular}
    \vspace{0.5em}
    \begin{tabular}{cc c|c|c|c c|c|c|c}
         &                                 & \multicolumn{3}{c}{$\langle W^e \rangle$ versus Quantum Discord} &                      &\multicolumn{3}{c}{$\Delta U_B$ versus Quantum Discord}&\\ \cline{3-5} \cline{7-9}
         &\multicolumn{1}{c|}{}            & \multicolumn{1}{c|}{\ \ \  Bi-partition \ \ \ }                 & \multicolumn{1}{c|}{\ \ \ \ \  Case 1 \ \ \ \ \ } & \multicolumn{1}{c|}{\ \ \ \ \  Case 2 \ \ \ \ \ }             & \multicolumn{1}{c|}{}& \multicolumn{1}{c|}{\ \ \  Bi-partition \ \ \ }          & \multicolumn{1}{c|}{\ \ \ \ \  Case 1 \ \ \ \ \ } & \multicolumn{1}{c|}{\ \ \ \ \  Case 2 \ \ \ \ \ }&\\  \cline{3-5} \cline{7-9}
         &\multicolumn{1}{c|}{Two-Qubit}   & $(A:B)$  & 0.05   & 0.03               & \multicolumn{1}{c|}{}& $(A:B)$  & -0.05  & -0.41 & \\ \cline{3-5} \cline{7-9}
         &\multicolumn{1}{c|}{Two-Qubit} & $(B:A)$ & 0.18   & -0.01               & \multicolumn{1}{c|}{}& $(B:A)$ & -0.18  & -0.81 & \\ \cline{3-5} \cline{7-9}
    \end{tabular}
    \caption{\label{tab:CCPrMaxWork}Pearson correlation coefficients for the two-qubit systems for the extractable work (left column) and $\Delta U_B$ (right-column). Top tables: Classical correlations; Bottom tables: Quantum discord.}
    
\end{table*}

\begin{figure}[hbt!]
    \centering
    \includegraphics[width=15.5cm]{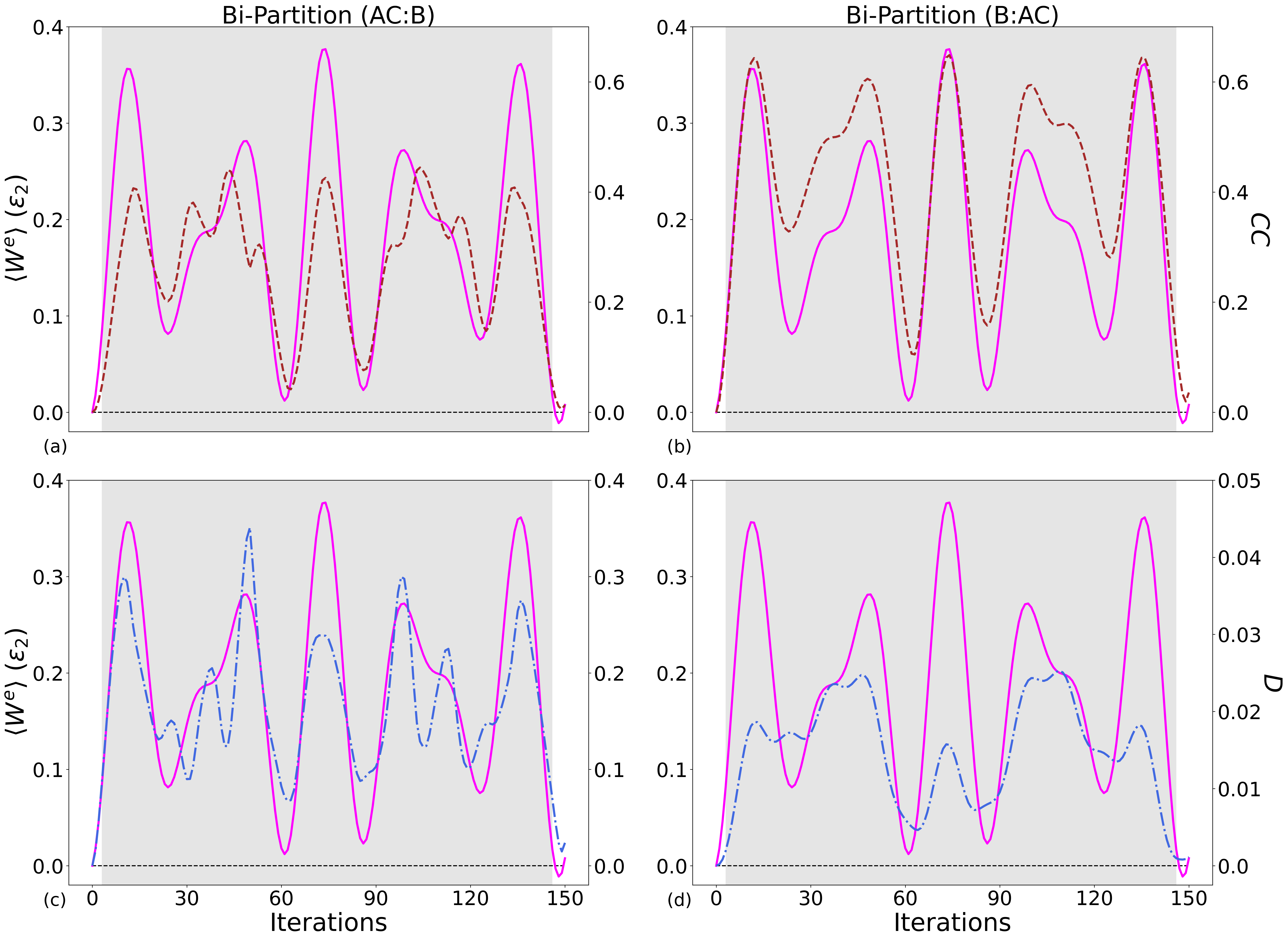}
    \caption{Same parameters as in Figure \ref{fig:C12BCCandD} but for the Case 1 three-qubit system.}
    \label{fig:C13BCCandD}
\end{figure}

\begin{table*}[hbt!]
\centering
    \begin{tabular}{cc c|c|c|c c|c|c|c}
         &                                 & \multicolumn{3}{c}{$\langle W^e \rangle$ versus Classical Correlations} &                      &\multicolumn{3}{c}{$\Delta U_B$ versus Classical Correlations}&\\ \cline{3-5} \cline{7-9}
         &\multicolumn{1}{c|}{}            & \multicolumn{1}{c|}{\ \ \  Bi-partition \ \ \ }                 & \multicolumn{1}{c|}{\ \ \ \ \  Case 1 \ \ \ \ \ } & \multicolumn{1}{c|}{\ \ \ \ \  Case 2 \ \ \ \ \ }             & \multicolumn{1}{c|}{}& \multicolumn{1}{c|}{\ \ \  Bi-partition \ \ \ }          & \multicolumn{1}{c|}{\ \ \ \ \  Case 1 \ \ \ \ \ } & \multicolumn{1}{c|}{\ \ \ \ \  Case 2 \ \ \ \ \ }&\\  \cline{3-5} \cline{7-9}
         &\multicolumn{1}{c|}{Three-Qubit}   & $(AC:B)$  & 0.81   & 0.25               & \multicolumn{1}{c|}{}& $(AC:B)$  & -0.81  & -0.47 & \\ \cline{3-5} \cline{7-9}
         &\multicolumn{1}{c|}{Three-Qubit} & $(B:AC)$ & 0.92   & 0.56               & \multicolumn{1}{c|}{}& $(B:AC)$ & -0.93  & -0.57 & \\ \cline{3-5} \cline{7-9}
    \end{tabular}

    \vspace{0.5em}
          
    \begin{tabular}{cc c|c|c|c c|c|c|c}
         &                                 & \multicolumn{3}{c}{$\langle W^e \rangle$ versus Quantum Discord} &                      &\multicolumn{3}{c}{$\Delta U_B$ versus Quantum Discord}&\\ \cline{3-5} \cline{7-9}
         &\multicolumn{1}{c|}{}            & \multicolumn{1}{c|}{\ \ \  Bi-partition \ \ \ }                 & \multicolumn{1}{c|}{\ \ \ \ \  Case 1 \ \ \ \ \ } & \multicolumn{1}{c|}{\ \ \ \ \  Case 2 \ \ \ \ \ }             & \multicolumn{1}{c|}{}& \multicolumn{1}{c|}{\ \ \  Bi-partition \ \ \ }          & \multicolumn{1}{c|}{\ \ \ \ \  Case 1 \ \ \ \ \ } & \multicolumn{1}{c|}{\ \ \ \ \  Case 2 \ \ \ \ \ }&\\  \cline{3-5} \cline{7-9}
         &\multicolumn{1}{c|}{Three-Qubit}   & $(AC:B)$  & 0.86   & 0.42               & \multicolumn{1}{c|}{}& $(AC:B)$  & -0.47  & -0.23 & \\ \cline{3-5} \cline{7-9}
         &\multicolumn{1}{c|}{Three-Qubit} & $(B:AC)$ & 0.57   & 0.16               & \multicolumn{1}{c|}{}& $(B:AC)$ & -0.57  & -0.29 & \\ \cline{3-5} \cline{7-9}
    \end{tabular}
    \caption{\label{tab:3BCCPrMaxWork}Pearson correlation coefficients for the three-qubit systems for the extractable work (left column) and $\Delta U_B$ (right-column). Top tables: Classical correlations; Bottom tables: Quantum discord.}
    
\end{table*}

Tables \ref{tab:CCPrMaxWork} and \ref{tab:3BCCPrMaxWork} contain the PCCs connecting work (left tables) and $\Delta U_{B}$ (right tables) to the classical correlations and quantum discord for all qubit systems and Cases combinations. 
Given the bi-partition $(B:A)$ for two-qubits and $(B:AC)$ for three, classical correlations correlate to $\langle W^e \rangle$ equally or stronger than the mutual information. This can be observed visually by comparing the upper panels of Figures \ref{fig:C12BCCandD} and \ref{fig:C13BCCandD} to results in Figure \ref{fig:MIMaxWorkSlices}. This shows that in general, when we generate work, there is a proportional creation of classical correlations between the components of our working medium. 

As for the mutual information, Case 2 three-qubit system is showing relatively low PCC's and for both bi-partitions: while some agreement is seen between the peaks of $\langle W^e \rangle$ and classical correlations, there is no consistency. 

In contrast to the mutual information and classical correlations no strong linear correlation is seen between the work, $\Delta U_B$ and the quantum discord, with the exception of Case 1 three-qubit system, PCC=0.86 with bipartition $(AC:B)$, as shown in the bottom left of Figure \ref{fig:C13BCCandD} and bottom table of Table \ref{tab:3BCCPrMaxWork}.  Here, $D(AC:B)$ is an order of magnitude greater than  $D(B:AC)$, meaning that the information that we can infer about qubit-B from quantum correlations attained by measuring qubits A and C is relatively small. As this is the only system where the quantum discord shows a strong linear correlation with $\langle W^e \rangle$ then this is not a universal feature of our systems performance but system and protocol specific. For Case 2, the quantum discord has the same order of magnitude than the classical correlations. Related figures are included in the Appendix \ref{appendF}. 

In summary, the mutual information is a strong indicator of when our system is producing larger amounts of $\langle W^e \rangle$, with work producing parameter regions generating correlations in the system with classical correlations playing a leading role. Classical correlations have equal if not better values for the PCC showing a strong linear correlation with the $\langle W^e \rangle$. This is, as expected, mirrored in the PCCs for $\Delta U_B$ which is generally the main component driving the production of work. 
Quantum discord does not show the same strength of linear correlation with $\langle W^e \rangle$ as either its classical counterpart or the mutual information, with the exception of the Case 1 three-qubit system and being zero for bipartition $(B:A)$. This suggests that, for this type of protocols, the generation and consumption of quantum correlations is in general poorly related to work production, being protocol and system size specific.

\section{Discussion}

Here we have shown that initial quantum coherence, solely provided by qubit-B's initialisation in a pure state, is advantageous to the production of extractable work. Indeed, for three out of four Case and system size combinations it increases the maximum work production (up to $28.20\%$) with respect to the initialization with all thermal qubits. The only Case that did not show improvement demonstrated an equal performance.  For all combinations, initial quantum coherences also extend the regions of extractable work to  $\theta \leq 0.5\pi$. These regions show a marked dependence on the azimuthal angle $\phi$.

As for any heat engine, classical or quantum, it is desirable to obtain a high efficiency when the engine is producing maximum work. All Cases and system sizes demonstrate a high efficiency at peak work with Case 1 protocols displaying an efficiency always at or close to unity and Case 2 protocols displaying an efficiency tracking more closely the peaks and troughs of work production. 

Our results also show the potential advantages of using NRCGs over full CNOT gates in work production. 
With the trotterized dynamics they induces, NRCG gates provide a more finely tuned and controllable work output for a quantum gate driven heat engine.
In addition, depending on protocol and iterations, NRCGs may lead to higher values of extractable work. Some related results are contained in Appendix \ref{appendC}.

The role of quantum correlations in thermal engines is an intricate subject, in part because it is somewhat model-dependent as discussed by Latune \textit{et al}.~\cite{Latune_2021}. Typically, related works are focused on advantages gained from, initial correlations~\cite{Herrera2023}, collective behaviour between $N$ identical copies of a heat engine~\cite{PhysRevLett.128.180602} or coherent baths~\cite{Hammam_2022}. In this work we have checked if a relation could be established between the extractable work production and three types of correlation: mutual information, classical correlation, and quantum discord. In our working medium these are solely generated through the application of our gate-based protocols. To quantify these relationships we have used the Pearson correlation coefficient.  We find a strong linear correlation between the extractable work and, both, the mutual information and classical correlations. For the latter, the partitions (B:A) and (B:AC) should be considered, with higher linear correlations for cases with lower values of corresponding quantum discord. This, together with other findings, suggests that the production of work is mainly tied to the production of classical correlations for this type of protocols.   

We also find, though outside of the scope of the work presented here, that the proposed system can operate in other regimes depending on initialisation parameters, such as a refrigerator or a reverse flow heat engine which is a uniquely quantum case. This naturally leads us towards further investigations and shows potential for the system to be used as a somewhat universal engine, with operating regime dictated by the needs of the operator. Further investigations could also include the addition of initial correlations between the components of the working medium and correlation between the working medium and the baths in the initialisation step.

\section{Conclusion}
     
In this paper we have shown that circuits utilising Nth-root CNOT gates can be advantageously used for quantum heat engines. We evaluate the systems performance by analysing the maximum work produced when one of the qubits, qubit-B, is initialised in a pure state and find this to be highly advantageous with respect to a system with initial thermal qubits and the same energy. Initial coherences not only improved maximum work production but also increased the size of work-producing parameter regions, extending them to areas corresponding to initial polar angle of qubit-B smaller than $\pi/2$. We also find that the protocols operates at very high to high efficiency when producing their maximum work, further demonstrating its potential use as a heat engine. Finally, we show that correlations described by the quantum mutual information between the components of the working medium demonstrate a strong linear correlation with the extractable work. It is similar for classical correlations when considering a specific partition order, rendering them a key component of work production. The systems and protocols we use are experimentally feasible with current technology.   
\vspace{6pt}

\noindent Conceptualization, E.F., I.D., T.M and F.S.; methodology, E.F. and I.D; software, E.F.; validation, E.F. and I.D.; formal analysis, E.F. and I.D.; investigation, E.F.; resources, I.D.; data curation, E.F.; writing---original draft preparation, E.F. and I.D.; writing---review and editing, E.F., I.D., F.S and T.M.; visualization, E.F.; supervision, I.D.; project administration, I.D.; funding acquisition, I.D. All authors have read and agreed to the published version of the manuscript. \\

\noindent EJF acknowledges support from EPSRC, grant number is EP/T518025/1; TMM thanks the grants No. 2021/01277-2 and No. 2022/09219-4, S\~{a}o Paulo Research Foundation (FAPESP); FSK acknowledges funding by the DFG within FOR 2724. \\
 
\noindent The authors declare no conflicts of interest.

\appendix
\section[\appendixname~\thesection]{Maximum extractable work when qubit-B is initialised in a thermal state}\label{appendB}

\begin{figure}[hbt!]
    \centering
    \includegraphics[width=15.5cm]{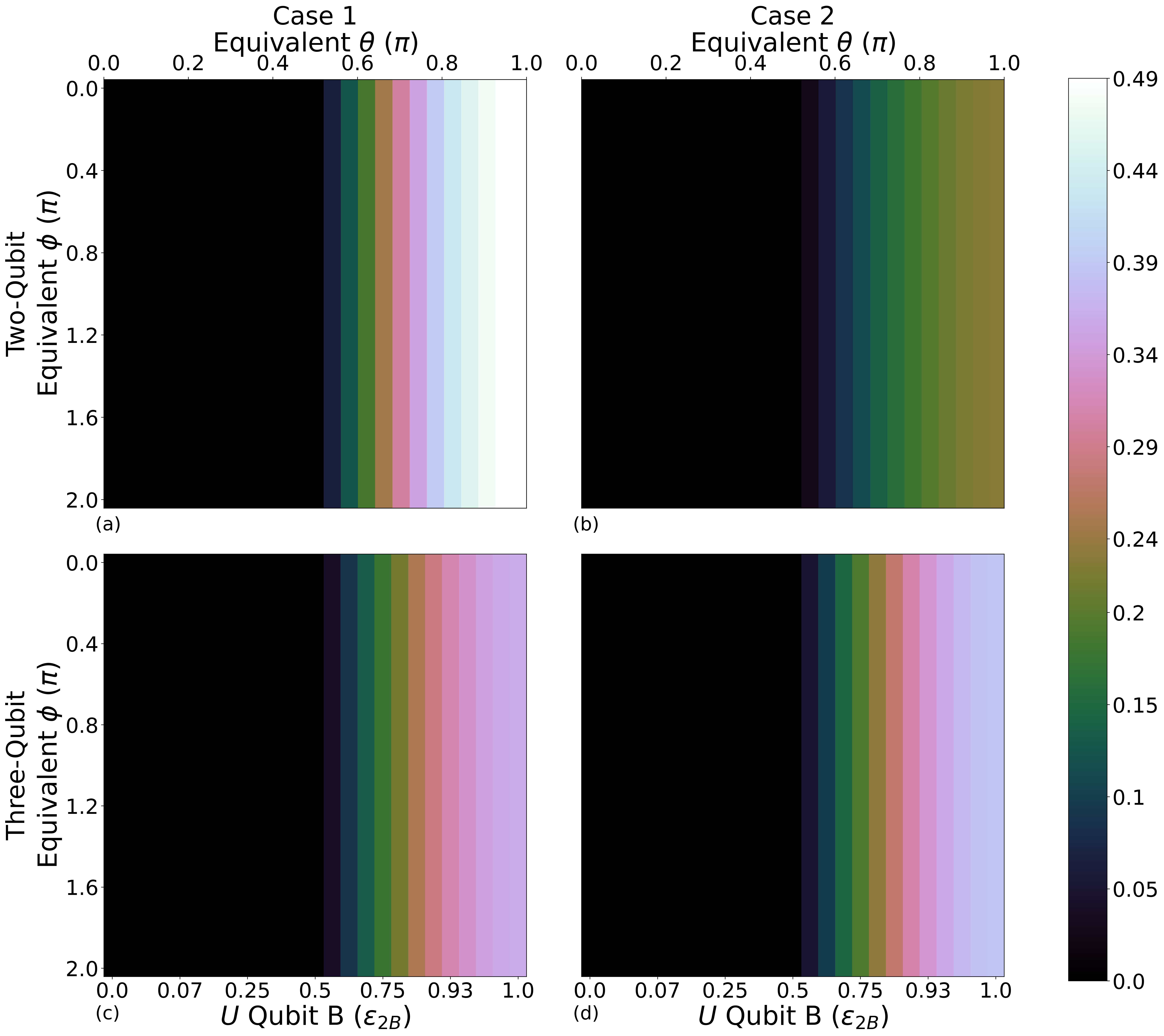}
    \caption{Maximum value of extractable work $\langle W^e_{max}\rangle$ in units of $\epsilon_{2}$ for the initial energy of qubit-B $0 \leq U^{(0)} \leq 1$ in units of $\epsilon_{2}$ (x-axis) and corresponding $0 \leq \phi \leq 2\pi$ (y-axis); First row: Two-qubit systems where columns left to right are cases 1 and 2 respectively. Brighter shades correspond to a greater maximum value of extractable work. Each combination of initial $U^{(0)}$ and $\phi$ is run for 150 iterations. For qubits A and C, $k_b\bar{T} = 40\epsilon_{2}$. Second row: The same parameters as the first row but for the three-qubit systems. All initial density matrices have no off-diagonal elements, all qubits being initialized in a Gibbs state.}    
    \label{fig:Completely thermal maximum work}
\end{figure}

\begin{table*}[hbt!]
\centering
    \begin{tabular}{c cc|c|c}
         &           &                       \multicolumn{2}{c}{$\langle W^e_{max} \rangle$} &\\ \cline{3-4} 
         &\multicolumn{1}{c|}{}   & \multicolumn{1}{c|}{\ \ \  Two-Qubit \ \ \ } & \multicolumn{1}{c|}{\ \ \  Three-Qubit \ \ \ } &\\ \cline{3-4} 
         &\multicolumn{1}{c|}{Case 1} & $0.49\epsilon_{2}$   & $0.36\epsilon_{2}$   &\\ \cline{3-4} 
         &\multicolumn{1}{c|}{Case 2} & $0.23\epsilon_{2}$   & $0.39\epsilon_{2}$   &\\ \cline{3-4} 
    \end{tabular}
    \caption{\label{table:thermal maximum work}Maximum values of extractable work when all component qubits are initialised in a thermal state.}
\end{table*}

In Figure \ref{fig:Completely thermal maximum work} qubit-B is initialised in a Gibbs state having the same energy $U$ (lower x-axis, non-linear scale) as the corresponding pure state characterized by the polar angle $\theta$ indicated in the upper x-axis (linear scale).
This allows for a direct comparison with Figure \ref{fig:Maxworkscansphi} where qubit-B is initialised in a pure state. 
As the off diagonal elements of a thermal state are zero, in  Figure \ref{fig:Completely thermal maximum work} there is no dependence of $\langle W^e_{max}\rangle$ on $\phi$.

There is an increase in maximum extractable work from when the equivalent $\theta$ increases from $0.5$ to $\pi$ for all Cases and system sizes, demonstrating the requirement for a degree of population inversion for qubit-B. The maximum extractable work is shown in Table \ref{table:thermal maximum work} and is compared and discussed in the main text.

\section[\appendixname~\thesection]{Comparison with circuits containing full CNOT gates}\label{appendC}

\begin{figure}[hbt!]
    \centering
    \includegraphics[width=15.5cm]{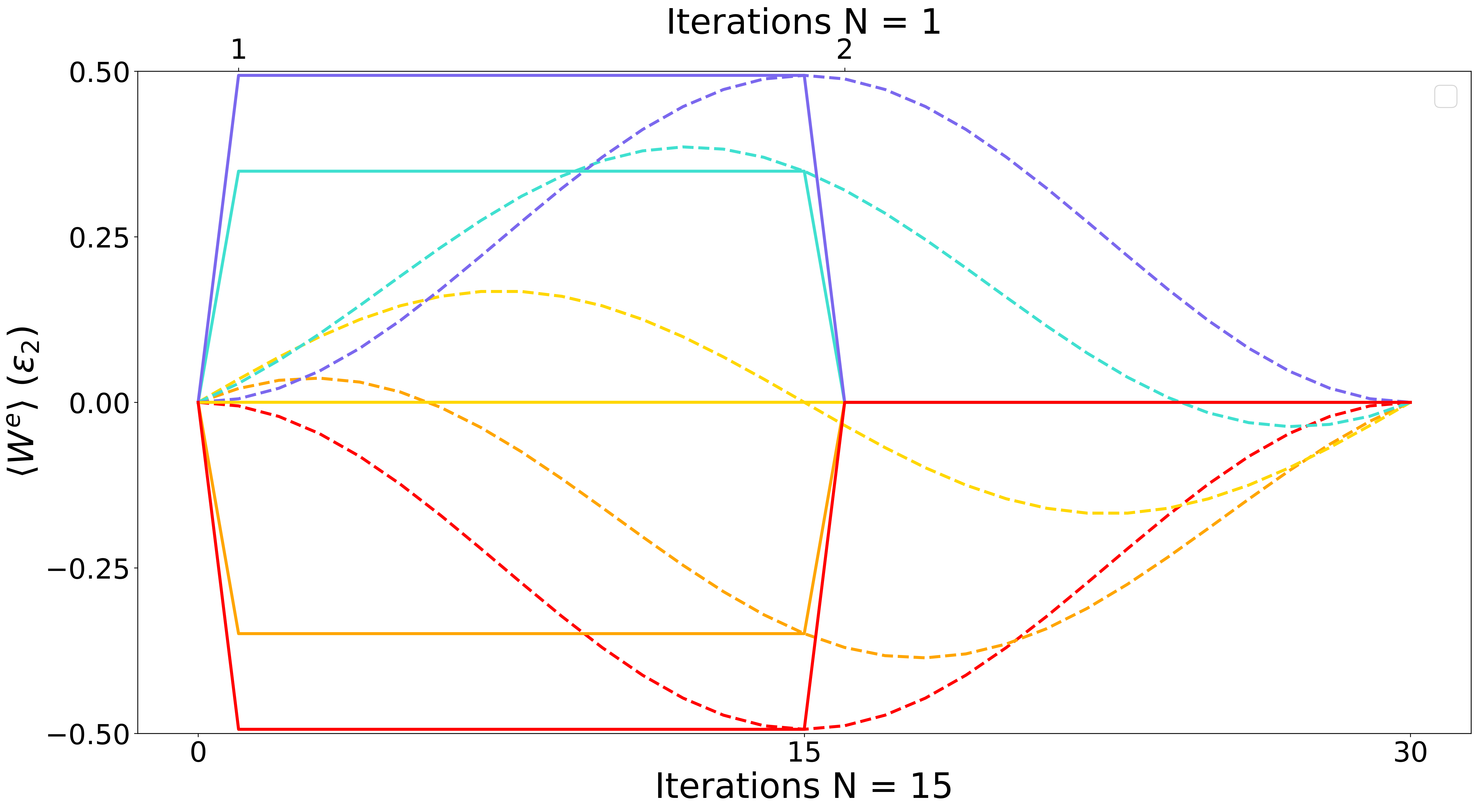}
    \caption{Comparing the extractable work $\langle W^e\rangle$ in units of $\epsilon_{2}$ (y-axis) of N = 15 (NRCG) and N = 1 (full CNOT gate) for the case 1 two-qubit system for four cycles. Same parameters as in Figure \ref{fig:Maxworkscansphi}. $\phi = 0.17\pi$. Red: $\theta = 0$; Orange: $\theta = 0.25\pi$; Yellow: $\theta = 0.50\pi$; Cyan: $\theta = 0.75\pi$; Blue: $\theta = \pi$. Dashed lines: N = 15; Solid lines: N = 1.}    
    \label{fig:varyingthetacnot1comparrisons}
\end{figure}

Figure \ref{fig:varyingthetacnot1comparrisons} compares $\langle W^e\rangle$ between an NRCG and a full standard CNOT gate for the Case 1 two-qubit system with varying $\theta$. As only 1 gate is applied in the protocol at each multiple of N iterations (a cycle) both gate types are equivalent. This can be seen as $\langle W^e\rangle$ is the same at the end of each cycle, regardless of the initial choice of $\theta$. When $ 0 < \theta < \pi$ we start to see where an advantage can be achieved. Within this range peaks and troughs of $\langle W^e\rangle$ for the NRCG are now not aligned with the peaks of the CNOT and are generally advantageous. We see unique behaviour for when $\theta$ is initialised at $0.50\pi$ (yellow). The full CNOT does not produce any work which is in contrast to the NRCG where we do see work production demonstrating the value of being able to access the intermediate states not readily accessible by a full CNOT. Comparing $\theta = 0.5\pi$ to the completely thermal initialisation in Figure \ref{fig:Completely thermal maximum work} where there is also no initial coherence, $\langle W^e_{max}\rangle = 0\epsilon_{2}$. The thermal and pure state initialisations of qubit-B are equivalent in initial energy and population of their ground and excited states. The only difference is the initial coherence, with non-zero off diagonal elements. Indicating initial coherence from the non-thermal initialisation coupled with the application of the NRCG is what provides this advantage over the full CNOT gates.

When $\theta = 0$ or $\theta = \pi$ any advantages attributed to peak work production disappear. For these values there is indeed no initial coherence.

\begin{figure}[hbt!]
    \centering
    \includegraphics[width=15.5cm]{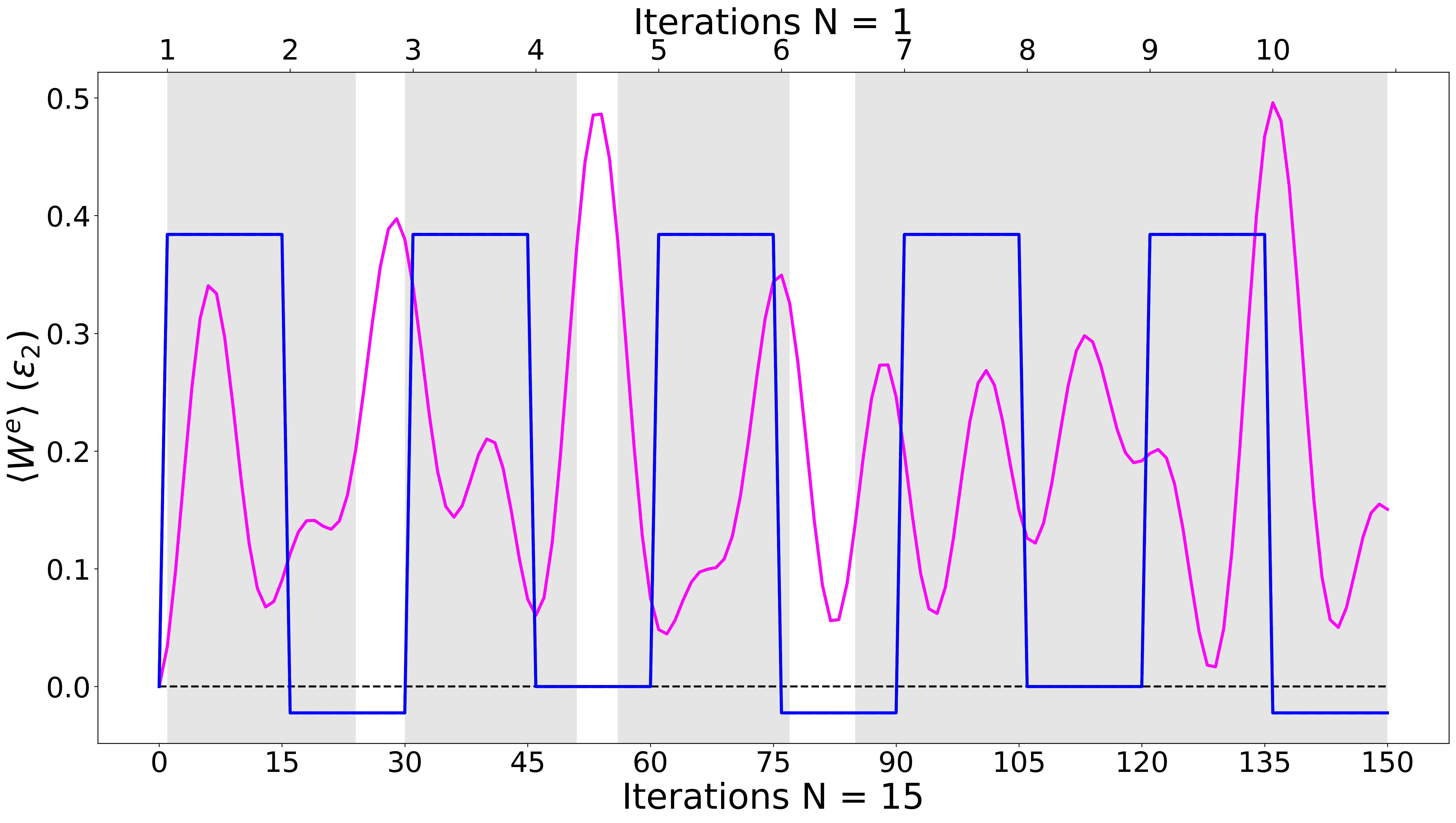}
    \caption{Comparing the extractable work $\langle W^e\rangle$ in units of $\epsilon_{2}$ (y-axis) for the NRCG-protocol with $N=15$ (magenta), and with the standard CNOT ($N=1$, blue), for Case 2, three-qubit systems. Bottom x-axis: $0 \leq$ iterations $\leq 150$ for $N=15$; Top x-axis: $0 \leq$ iterations $\leq 10$ for $N=1$. Same parameters as in Figure \ref{fig:Maxworkscans}.}    
    \label{fig:Cnot Comparrisons}
\end{figure}

Figure \ref{fig:Cnot Comparrisons} shows a comparison of circuits built with NRCGs and $N=15$ against circuits with full standard CNOT gates (i.e. $N=1$) when the Case 2 three-qubit system is initialised for maximum work production, as in Figure \ref{fig:Maxworkscans}. Here the grey shaded regions correspond to when the NRCG protocol (magenta line) is operating in the heat engine regime. The protocol is applied for $M = 10N$ iterations for both gate types: 150 iterations when N = 15; 10 iterations when N = 1.  Circuits with NRCGs show larger peaks of work production than the ones with full CNOT gates at 30, 55, and 137 iterations, though only the peak at 137 iterations falls within the heat engine regime. The NRCG-based circuit always produces work, while in same regions work must be provided to the system for CNOT-based protocols, the corresponding $\langle W^e\rangle$ being less than zero.
In general, the use of NRCGs demonstrates more control over the quantity of extractable work, exploiting intermediate states not accessible by full CNOT gates. This feature provides a benefit for the quantum heat engine: as the choice of when to halt the application of the protocol resides with the user, an iteration can by chosen that corresponds to the desired value of $\langle W^e \rangle$. 

\section[\appendixname~\thesection]{Derivation of the quantum discord from the mutual information for the one, and two-qubit projective measurements}\label{appendD}

Here, we give the full derivation of the quantum discord from the mutual information for both the one, and two-qubit projective measurements. The mutual information captures all information shared between the components of a chosen bi-partition encapsulating both classical and quantum correlations. The mutual information has two equivalent classical expressions,

\begin{equation}
        I(A:B) = H(A) + H(B) - H(AB),
    \label{eq:classicalMIsapend}
\end{equation}
    and
\begin{equation}
    J(A:B) = H(A) - H(A|B) = H(B) - H(B|A).
    \label{eq:classicalMIsconditionalapend}
\end{equation}

The quantum generalisation for the first expression of the mutual information quantifies the total correlations in some bi-partite system which includes both quantum and classical parts \cite{PhysRevA.71.062307}. It is symmetric about the bi-partition and has the form, 

\begin{equation}
    I(\rho_{AB}) = S(\rho_A) + S(\rho_B) - S(\rho_{AB}).
    \label{eq:mutualinformationapend}
\end{equation}

Here the Shannon entropy $H(A)$ \cite{cover1999elements} has been replaced with the von Neumann entropy \cite{Nielsen_Chuang_2010,Peres1995-PEREQT} which is equivalent when using this expression for quantum systems. The von Neumann entropy has the form,

\begin{equation}
    H(A) = S(\rho_A) = -tr[\rho_A \ln{\rho_A}].
\end{equation}

The second term in eq \ref{eq:classicalMIsconditionalapend} requires determining the conditional state of subsystem A. The quantum description of $J(A:B)$ is interpreted as the information gained about subsystem A with a measurement on subsystem B and has the form,   

\begin{equation}
    J(\rho_{AB})_{\{\Pi_j^B \}} = S(\rho_A) - S(\rho_A|\{\Pi_j^B \}).
    \label{eq:classicalcorrelationsapend}
\end{equation}

where $S(\rho_A|\{\Pi_j^B \})$ is the conditional entropy.  Maximisation over $J(\rho_{AB})$ encompasses the maximum classical correlations that can be obtained from the chosen bi-partition with any remaining correlations being completely quantum. Hence $S(\rho_{A}| \{ \Pi^B_j \})$ is maximised over the complete orthonormal measurement basis $\{\ket{u}, \ket{v}\}$ and is given by

\begin{equation}
    S(\rho_{A}| \{ \Pi^B_j \}) = \sum_{j = u, v}  p_j S(\rho_{A| \Pi^B_j}),
\end{equation}  

\noindent which is the sum of the von Neumann entropies of the post measurement states over the complete set of projective measurements. This yields  the conditional entropy of A given the complete measurement $\{\Pi^B_j\}$ on B. The complete orthonormal measurement basis $\{ \ket{u}, \ket{v} \}$ is, 

\begin{equation}
    \{ \ket{u} = \cos{(\theta/2)} \ket{0} + \sin{(\theta/2)}e^{i\phi}\ket{1}, \ket{v} = \sin{(\theta/2)}e^{-i\phi}\ket{0} - \cos{(\theta/2)} \ket{1}\}.
\end{equation}

This basis is used to create projectors for the measurement $\Pi^B_u = I^A \otimes \ket{u}\bra{u}$ and $\Pi^B_v = I^A \otimes \ket{v}\bra{v}$ where $I^A$ is the identity matrix with dimensionality corresponding to that of subsystem A. The post measurement state of A that corresponds to the outcome of the measurement on B has the form,

\begin{equation}
    \begin{gathered}
        \rho_{A| \Pi^B_j} = ( \Pi^B_j) \rho_{AB} ( \Pi^B_j) / p_j, \ \ p_j = tr[\Pi^B_j 
        \rho_{AB}]. 
    \end{gathered}
\end{equation}

 With a quantum equivalent now defined for both expressions of mutual information we can now define the quantum discord. The discord has the form,

\begin{equation}
    \begin{gathered}
    D(A:B)_{\{\Pi_j^B \}} = I(A:B) - \max_{\Pi^B} J(A:B)_{\{\Pi_j^B \}},\\ 
    D(A:B)_{\{\Pi_j^B \}} = \min_{\Pi^B}[ S(B) - S(AB) + S(A|\{ \Pi ^B_j\})].    
    \label{eq:discordapend}    
    \end{gathered}
\end{equation}

To satisfy the minimisation criteria the discord is calculated over $0 > \theta > \pi$ and $0 > \phi > 2\pi$. 
Quantum discord for the three-qubit system requires an expansion of our measurement basis to cover two qubits simultaneously due to the bi-partition in one and two qubits. This measurement basis is

\begin{equation}
    \{ \ket{u}\ket{u'},\ket{u}\ket{v'},\ket{v}\ket{u'},\ket{v}\ket{v'} \},
\end{equation}

where

\begin{equation}
\begin{gathered}
    \ket{u} = \cos{(\theta/2)} \ket{0} + \sin{(\theta/2)}e^{i\phi}\ket{1}, \ket{v} = \sin{(\theta/2)}e^{-i\phi}\ket{0} - \cos{(\theta/2)} \ket{1} \\
    \ket{u'} = \cos{(\theta'/2)} \ket{0} + \sin{(\theta'/2)}e^{i\phi'}\ket{1}, \ket{v'} = \sin{(\theta'/2)}e^{-i\phi'}\ket{0} - \cos{(\theta'/2)} \ket{1}.
\end{gathered}
\end{equation}

Projectors are created in the same way as for the single qubit measurement,

\begin{equation}
\begin{gathered}
    \Pi^{AC}_{uu'} = \ket{u}\bra{u} \otimes I \otimes \ket{u'}\bra{u'}, \ \Pi^{AC}_{vu'} = \ket{v}\bra{v} \otimes I \otimes \ket{u'}\bra{u'}, \\
    \Pi^{AC}_{uv'} = \ket{u}\bra{u} \otimes I \otimes \ket{v'}\bra{v'}, \ \Pi^{AC}_{vv'} = \ket{v}\bra{v} \otimes I \otimes \ket{v'}\bra{v'}.
\end{gathered}
\end{equation}

\noindent Measurement is performed for all values of $0 \leq \theta \leq \pi$, $0 \leq  \phi \leq 2\pi$, $0 \leq  \theta' \leq \pi$, and $0 \leq \phi \leq 2\pi'$.

\section[\appendixname~\thesection]{Classical correlations and quantum discord for Case 2}\label{appendF}

\begin{figure}[hbt!]
    \centering
    \includegraphics[width=15.5cm]{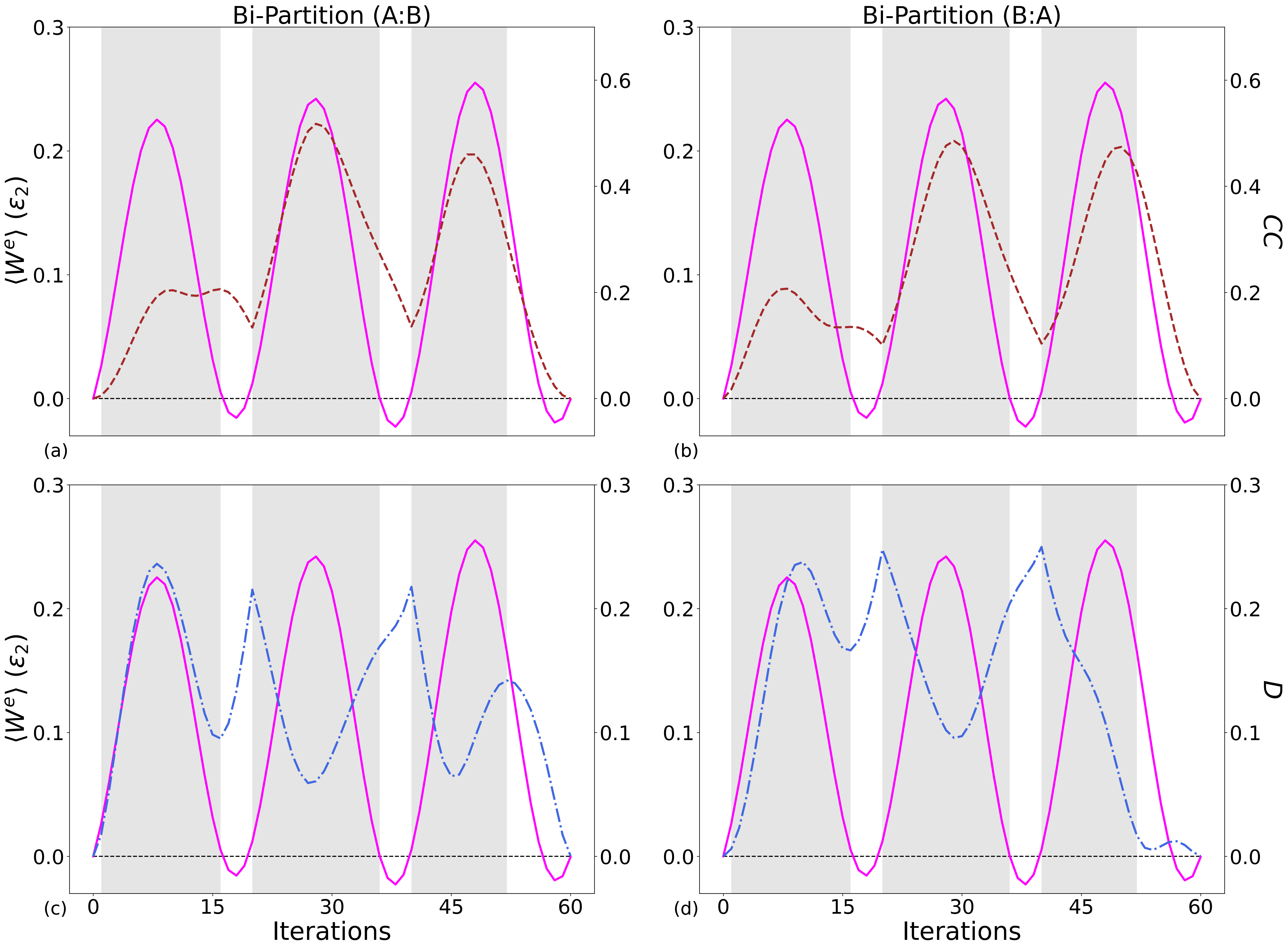}
    \caption{Same parameters as in Figure \ref{fig:C12BCCandD} but for the Case 2 two-qubit system.}    
    \label{fig:C32BCCandD}
\end{figure}

The classical correlations and quantum discord for the Case 2 two-qubit system are shown in Figure~\ref{fig:C32BCCandD}, the quantum discord is non analytic at 21 and 41 iterations, abruptly reducing in value showing a cusp. At these iterations we observe the system re-entering the heat engine regime which is signified here by $\langle W^e \rangle < 0$ evolving to $\langle W^e \rangle > 0$. This is the sole change occurring at these iterations which removes this protocol and system size from the heat engine regime, suggesting that for this Case as, by increasing iterations, we cross the work-producing threshold quantum correlations are consumed, accompanied by a sudden rise in classical correlations.

\begin{figure}[hbt!]
    \centering
    \includegraphics[width=15.5cm]{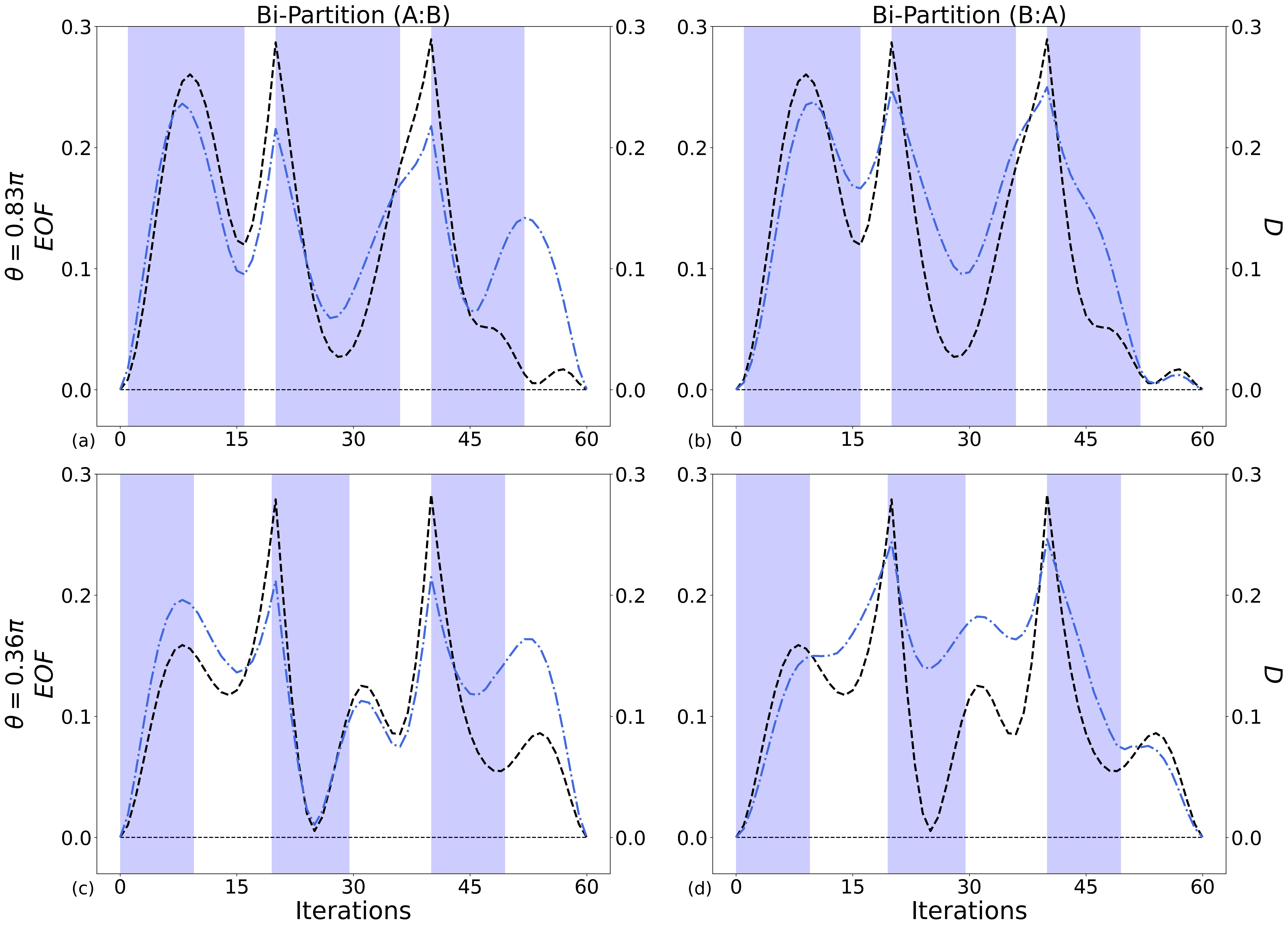}
    \caption{Comparison of the entanglement of formation (left y-axis) and quantum discord (right y-axis) for $0 \leq$ iterations $\leq 60$ (x-axis) for Case 2 two-qubit system. Top row: $\theta = 0.83\pi$; Bottom row: $\theta = 0.36\pi$. Blue shaded regions indicate work production, i.e regions with $\langle W^e \rangle > 0$. These include (but not coincide with) the regions in which the system behaves as a heat engine. Black dashed: Entanglement of formation; Blue dash dot: Quantum discord.}    
    \label{fig:EOF contributions}
\end{figure}

This property is mirrored in the entanglement of formation, see Figure \ref{fig:EOF contributions}, suggesting that it is the sudden reduction in entanglement driving a reduction in the quantum discord. This is not only seen for the parameters of Figure \ref{fig:C32BCCandD} (first row of Figure \ref{fig:EOF contributions}) but also for other initial $\theta$ for qubit-B (e.g. second row of Figure \ref{fig:EOF contributions}), suggesting this to be a consistent behaviour for the protocol. In comparison, for the Case 1 two-qubit or either of the three-qubit systems, no entanglement of formation is produced for any of the possible bi-partitions and no cusps are seen, rendering this a unique feature.  

\begin{table*}[hbt!]
\centering
    \begin{tabular}{c cc|c|c}
         &&\multicolumn{2}{c}{Entanglement of Formation versus Quantum Discord} &\\ \cline{3-4} 
         &\multicolumn{1}{c|}{} & \multicolumn{1}{c|}{\ \ \ \ \  Bi-Partition $(A:B)$ \ \ \ \ \ } & \multicolumn{1}{c|}{\ \ \ \ \  Bi-Partition $(B:A)$ \ \ \ \ \ } &\\ \cline{3-4} 
         &\multicolumn{1}{c|}{$\theta = 0.83\pi$} & 0.85   & 0.91   &\\ \cline{3-4} 
         &\multicolumn{1}{c|}{$\theta = 0.36\pi$} & 0.81 & 0.78   &\\ \cline{3-4} 
    \end{tabular}
    \caption{\label{tab:EOF contributions table}PCCs for the quantum discord and the entanglement of formation when the system operates in a maximum work producing regime and when we operate in the mode of work symmetry.}
\end{table*}

In Figure \ref{fig:EOF contributions}, we show the entanglement of formation and quantum discord for the Case 2 two-qubit system to emphasize the similarity in behaviour of the entanglement of formation and quantum discord discussed previously. The entanglement of formation has the form \cite{10.5555/2011326.2011329}

\begin{equation}
    EOF_{AB} = -xlog_2x-(1-x)log_2(1-x),
\end{equation}

where $x=\frac{1-\sqrt{1-\tau}}{2}$, $\tau = [max(\lambda_1-\lambda_2-\lambda_3-\lambda_4, 0)]$, $\lambda_i = \sqrt{\epsilon_i}$, and $\epsilon_i$ are the eigenvalues of the matrix $\rho_{AB}\overline{\rho_{AB}}$. Here $\rho_{AB}$ is the reduced density matrix of sites A and B and $\overline{\rho_{AB}}$ is the spin-flipped $\rho_{AB}$, so $\overline{\rho_{AB}}=(\sigma^A_y\otimes\sigma^B_y)\rho_{AB}^*(\sigma^A_y\otimes\sigma^B_y)$. The PCC values shown in Table~\ref{tab:EOF contributions table} demonstrate a strong linear correlation between the discord and the EOF.   

\begin{figure}[hbt!]
    \centering
    \includegraphics[width=15.5cm]{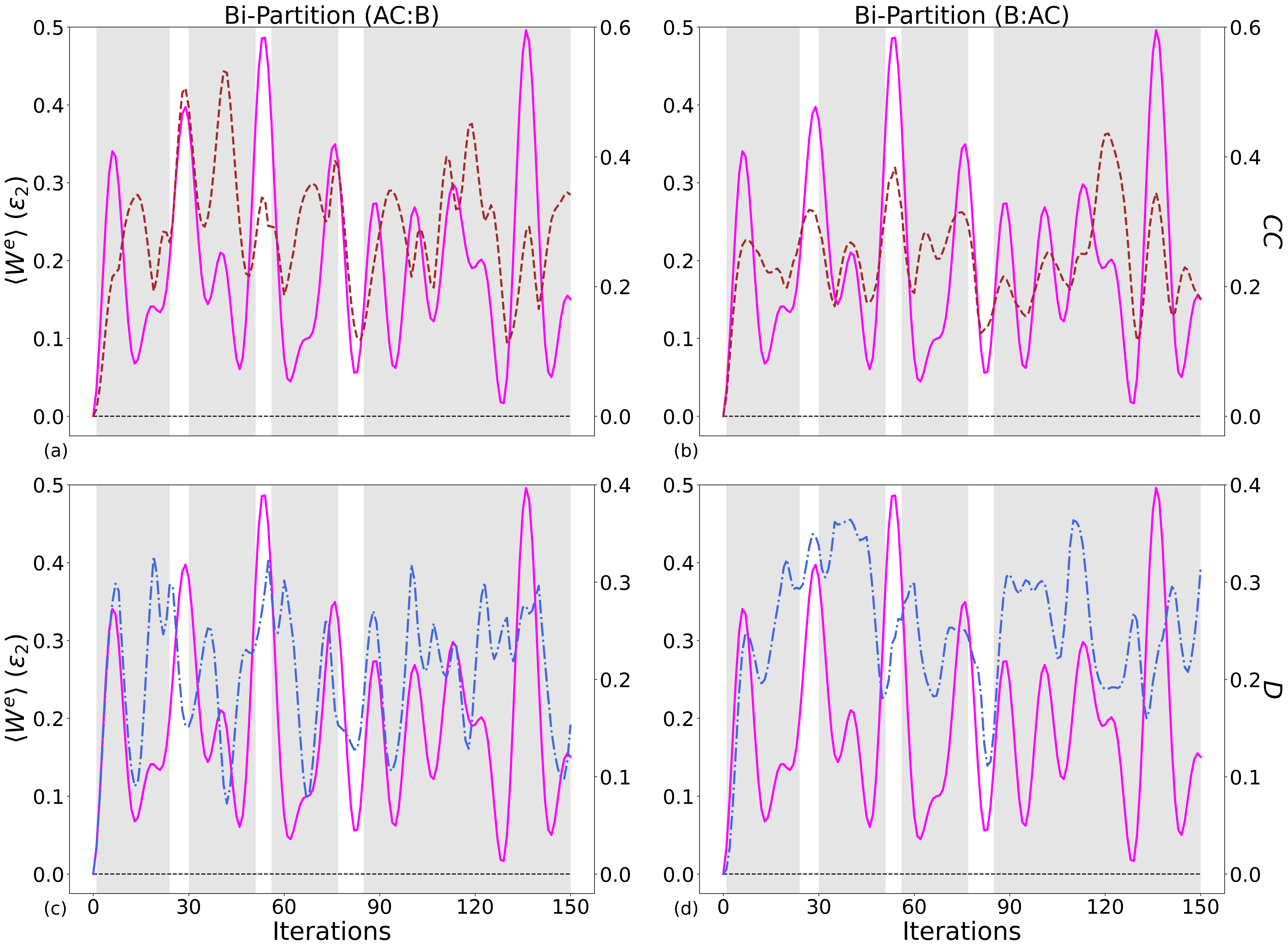}
    \caption{Same parameters as in Figure \ref{fig:C12BCCandD} for the Case 2 three-qubit system.}    
    \label{fig:C33BCCandD}
\end{figure}

Finally we examine the different correlation types for the Case 2 three-qubit system in Figure~\ref{fig:C33BCCandD}. It can be seen that the evolution with protocol application of the correlations is complex, being challenging to identify a relationship between the $\langle W^e \rangle$ and both the classical correlations and the quantum discord. Some peaks of the classical correlations correspond to peaks of the $\langle W^e \rangle$, this can be seen in the PCC value for the (B:AC) bi-partition (Table~\ref{tab:MIPrMaxWork}), being greater than that of the mutual information (Table~\ref{tab:3BCCPrMaxWork}). Even so, there is no clear observable  linear correlation between $\langle W^e \rangle$ and the two types of correlation that contribute to the mutual information.

\clearpage

\bibliography{library.bib}

\end{document}